\documentclass[a4paper,12pt]{article}

\usepackage[T2A]{fontenc}
\usepackage[cp1251]{inputenc}
\usepackage[russian]{babel}
\usepackage{graphicx}

\usepackage[centertags]{amsmath}
\usepackage{amsfonts}
\usepackage{amssymb}
\usepackage{amsmath}
\usepackage{newlfont}
\usepackage{caption}
\usepackage{longtable}
\usepackage{array}
\usepackage{lscape}

\newlength{\defbaselineskip}
\newcommand{\setlinespacing}[1]%
           {\setlength{\baselineskip}{#1 \defbaselineskip}}

\begin{document}

 \title{\bfseries\boldmath Тонкий эффект реакции $nd\to p(nn)$}
 \maketitle

\begin{center}
 \author{Р.\,А.~Шиндин$^*$, Д.\,К.~Гурьев, А.\,Н.~Ливанов, И.\,П.~Юдин}
 \vskip 3mm
 {\small{\it ОИЯИ, Лаборатория Физики Высоких Энергий, 141980 Дубна, Россия}\\
 $*\quad$ {\it E-mail: romanshindin@yandex.ru}}
\end{center}

\vskip 3mm

 \hfill Pacs: {25.40.Kv}

 \hfill UDC: {539.171.11}

\vskip 3mm



\begin{abstract}
 В процессе $nd\to p(nn)$ перезарядки нейтрона на дейтроне
 при рассеянии протонов вблизи нуля дейтрон переходит в пару нейтронов
 с малыми относительными импульсами.
 Чтобы объяснить особенности спектра вторичных протонов оказалось
 недостаточно учесть потери,
 связанные с компенсацией энергии связи
 $\varepsilon_{\textrm{св}}\approx2.23$\,МэВ дейтрона.
 Это приводит к предположению,
 что два нейтрона образуют компаунд-систему
 со своим собственным распределением по импульсам Ферми.
 \end{abstract}

 \maketitle


 \section{Введение}
 Основная задача эксперимента Дельта-Сигма --- полный опыт
 \cite{Lapidus-ppt-method,Lehar-ppt-method,Sharov-dsl-measurement,Sharov-dsl-measurements}
 восстановления трёх амплитуд $NN$-рассеяния под нулём при энергиях T$_n=1.2\div3.7$\,ГэВ.
 Чтобы исключить знаковую неоднозначность
 в процедуре прямого восстановлении амплитуд,
 начиная с 2002 г. мы провели измерения наблюдаемой $R_{dp}$
 --- отношения выходов вторичных протонов квазиупругой $nd\to p(nn)$
 и упругой $np\to pn$ реакций перезарядки под нулём градусов
 в диапазоне энергий T$_n=0.5\div2.0$\,ГэВ.
 В теории данное отношение выражается формулой Дина \cite{Dean-1}:
 \begin{equation}\label{introduction.Rdp-ratio.formula}
    R_{dp}(0)\;=\;
     \dfrac{d\sigma(0)}{dt}_{nd\to p(nn)}\;\Bigl/\;\dfrac{d\sigma(0)}{dt}_{np\to pn}
     = \;\frac{2}{3}\cdot\frac{1}{1+r^\textrm{nfl/fl}_{np\to pn\,(0)}}\quad,
  \end{equation}
 что позволяет детерминировать отношение $r^\textrm{nfl/fl}_{np\to pn\,(0)}$
 между spin-flip и spin-non-flip частями дифференциального сечения
 упругой $np\to pn$ перезарядки \cite{Shindin-Flip-PEPAN}.
 Импульсное приближение, в котором получена формула \eqref{introduction.Rdp-ratio.formula},
 позволяет считать нейтрон дейтрона спектатором
 и пренебречь его взаимодействием с налетающим нейтроном.
 На первый взгляд, неупругость реакции $nd\to p(nn)$ связана лишь
 с компенсацией энергии связи $\varepsilon_{\textrm{св}}\approx2.23$\,МэВ ядра дейтерия.
 Это было бы справедливо, если нейтрон спектатор оставить в покое,
 а весь переданный от рассеянного протона импульс $\vec{q}$ сообщить нейтрону отдачи,
 или в том случае, если предположить рождение динейтрона
 \cite{Migdal,Baz-Goldansky-Zeldovich} с массой $2m_n\approx1879$\,МэВ/$c^2$.
 Однако энергетические потери $\delta{E}=E_n-E_p$,
 где $E_{n}$ и $E_{p}$ ---  средние значения полных энергий нейтрона пучка и рассеянного протона,
 превосходят $\varepsilon_{\textrm{св}}$ в 2.5 раза,
 что говорит о возникновении промежуточного состояния $nn$-пары,
 инвариантная масса которого больше двух нейтронов.

\section{Установка Дельта-Сигма}\label{chapter-Setup}
\begin{figure}[!ht]
 \centering
 \scalebox{.14}{\includegraphics{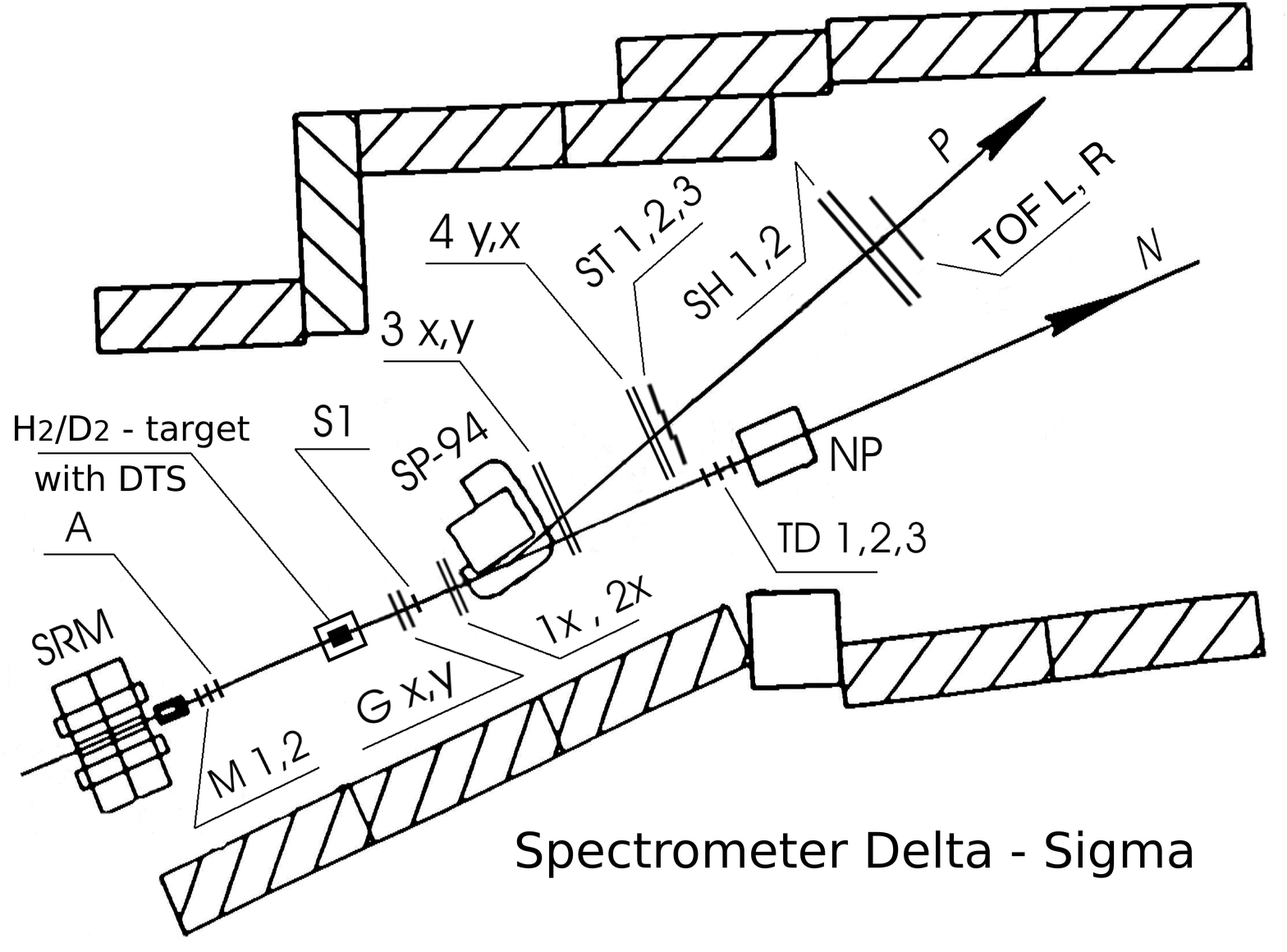}}
 \vspace {-5mm}
 \caption{\footnotesize Спектрометр для регистрации
 протонов упругой $n\to p$ перезарядки.
 SP-94 -- анализирующий магнит, Gxy, 1x, 2x, 3xy и 4xy --
 пропорциональные камеры,
 H$_2$/D$_2$ -- водородная или дейтериевая мишень,
 окружённая вето-системой ДОМ (DTS),
 M\,1,2 и TD\,1,2,3 -- мониторные и трансмиссионные детекторы нейтронов,
 A -- анти-счётчик,
 S1 и ST\,1,2,3 -- триггерные счётчики,
 SH\,1,2 -- сцинтилляционные годоскопы
 и времяпролётная система TOF = S1 + TOF\,L, R.}\label{setup.spectrometer}
 \end{figure}

 Для измерений импульсных
 и угловых распределений протонов реакции $n\to p$ перезарядки,
 используется спектрометр, схема которого показана на рис.~\ref{setup.spectrometer}.
 \noindent
 Пучок нейтронов диаметром $\varnothing\approx25$\,мм
 имеет расходимость $\sigma{\theta}\approx2$\,мрад
 и разброс импульсов $\sigma{P}/P\approx3.3$\,\%.
 Интенсивность пучка определятся мониторными детекторами нейтронов M1 и M2.
 Счётчик A исполняет роль вето для исключения фона заряженных частиц,
 рожденных на веществе конверторов мониторов.
 На пути нейтронов размещаются криогенные D$_2$/H$_2$-мишени
 или твёрдые CD$_2$/CH$_2$/C-мишени,
 комплементарные друг другу по количеству атомов дейтерия, водорода и углерода.
 В ходе реакции $n\to p$ перезарядки образуются вторичные протоны,
 треки которых до магнита СП-94 восстанавливают
 многопроволочные пропорциональные камеры Gxy, 1x и 2x.
 Треки частиц после их взаимодействия с магнитным полем определяются по камерам 3xy и 4xy.
 Триггер установки формируется совпадением сигналов от счётчика S1
 и блока счётчиков ST\,1,2,3.
 Импульсное и угловое разрешения спектрометра составляют
 $\sigma{P}/P\approx0.7$\,\% и $\sigma{\theta}\approx1.2$\,мрад.
 Направление вперёд ограничивается в пределах углов $\Delta{\theta}<50$\,мрад.
 Времяпролётная система TOF (Time-of-Flight)
 предназначена для исключения примеси дейтронов,
 появляющихся в реакции захвата $np\to d\pi^0$.
 Методику работы с этой системой раскрывает пункт \ref{chapter-Setup.tof}.
 Годоскоп сцинтилляционных счётчиков SH\,1,2 используется
 для контроля направления вторичных протонов
 на центр детектора TOF\,L, R.
 Одним из препятствий изучения спектров протонов при энергиях T$_n=1.4\div2.0$\,ГэВ
 стал неупругий фон, связанный в основном с возбуждением резонанса $\Delta$\,(1232).
 Для исключения таких событий используется детектор окружения мишени ДОМ (DTS),
 описанию которого посвящён пункт \ref{chapter-DTS}.

 \subsection{Времяпролётная система TOF}\label{chapter-Setup.tof}
\begin{figure}[!ht]
 \centering
 \scalebox{.11}{\includegraphics{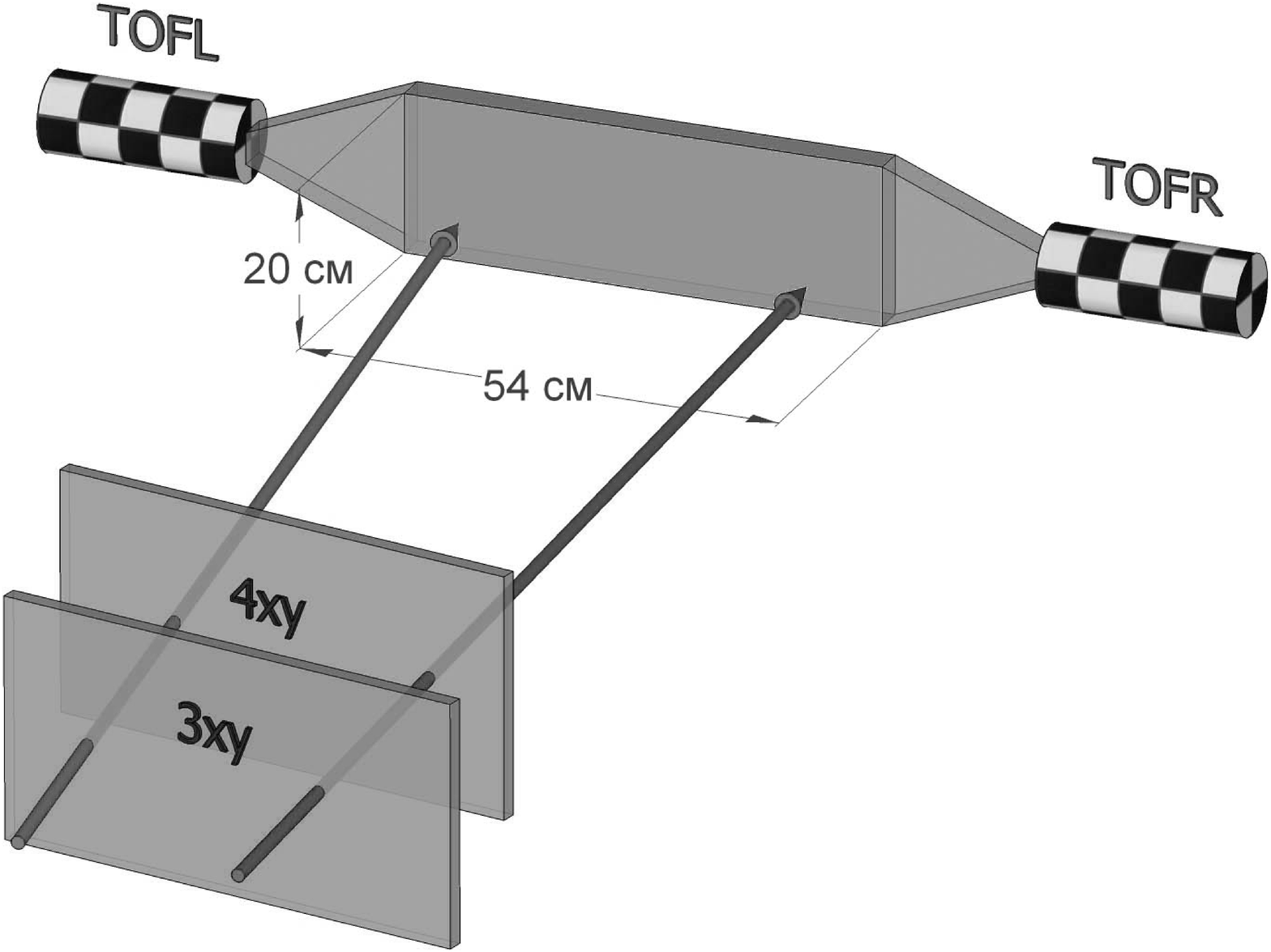}}\qquad
 \scalebox{.22}{\includegraphics{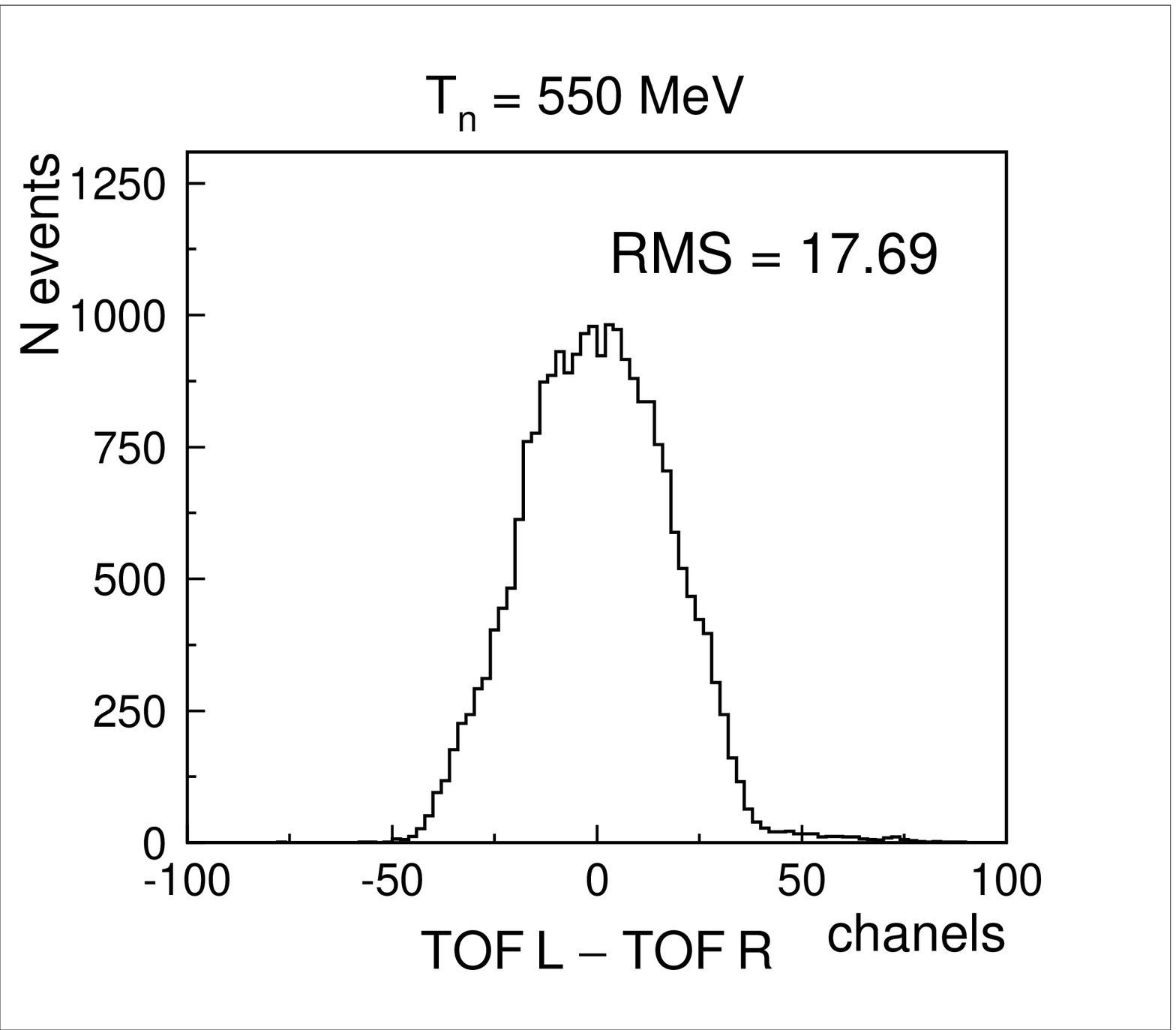}}
 \caption{\footnotesize На картинке слева представлен
 двухплечевой детектор TOF\,L, R,
 размещенный позади пропорциональных камер 3xy и 4xy на расстоянии $\sim7.7$ м.
 Левое TOF\,L и правое  TOF\,R плечо детектора формируют сигналы STOP времяпролётной системы,
 сигнал START которой берётся от триггерного счётчика S1
 (рис.~\ref{setup.spectrometer}).
 Стрелками схематично показаны треки частиц.
 Справа приведена гистограмма разности времён прихода сигналов левого и правого плеча.
 Разброс этой величины составляет 17.7 канала ВЦП, что эквивалентно 1.8\,нс.
 }\label{setup.tof-schem}
 \end{figure}

\begin{figure}[!ht]
 \centering
 \scalebox{.25}{\includegraphics{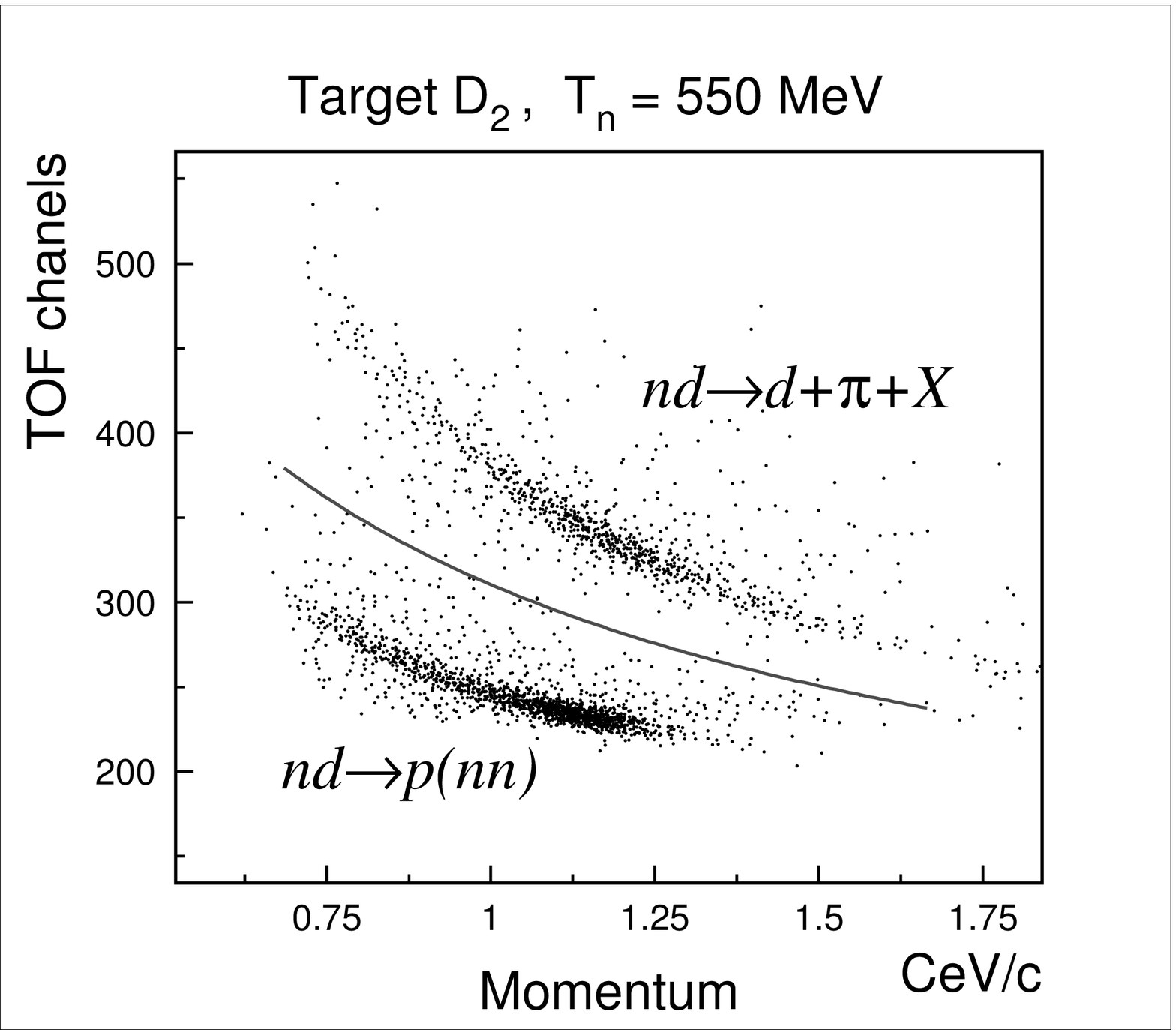}}\quad
 \scalebox{.25}{\includegraphics{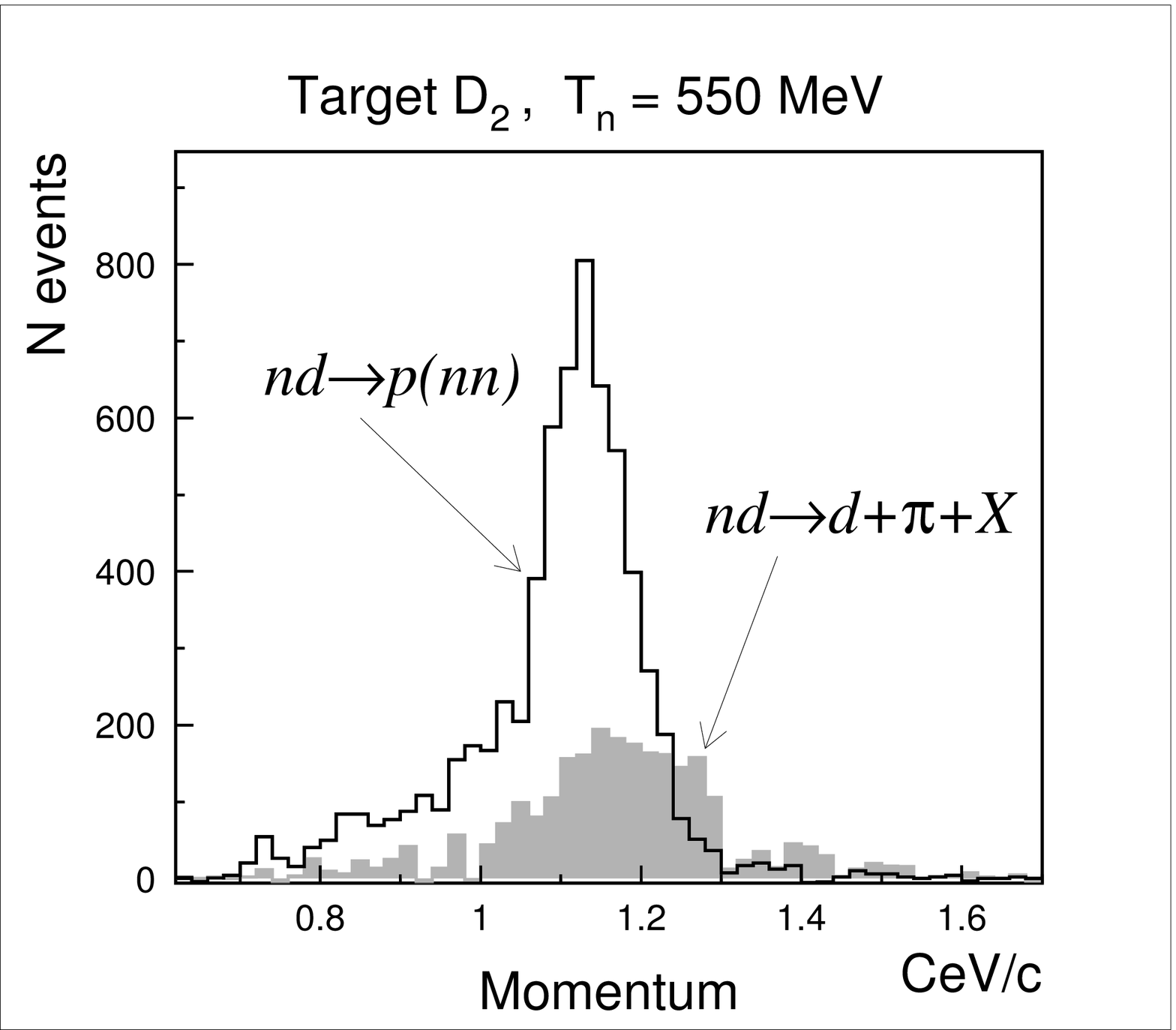}}
 \caption{\footnotesize Двумерный плот слева в координатах
 импульса и времени пролёта (1\,канал\,=\,100\,пс) представляет
 распределения реакций $nd\to p(nn)$ и $nd\to d+\pi+X$
 при энергии T$_n=550$\,МэВ и углах рассеяния, близких к нулю.
 Вторичные дейтроны и протоны разделяются гиперболой.
 Результаты такого способа обработки представлены на гистограмме справа.
 }\label{setup.tof2dimentions}
 \end{figure}

 TOF-схема включает в себя два детектора
 S1 (рис.~\ref{setup.spectrometer})
 и TOF\,L, R (рис.~\ref{setup.tof-schem}, слева),
 \noindent
 расположенных на базе $\sim10$\,м друг за другом. 
 Счётчик S1
 даёт сигнал начала отсчёта двум время-цифровым преобразователям TDC\,2228,
 цена канала в которых составляет $\sim100$\,пс.
 Когда частица попадает в детектор TOF\,L, R,
 его левое TOF\,L и правое TOF\,R плечо формируют свои стоповые сигналы.
 За счёт разности хода света в сцинтилляторе
 TOF\,L и R срабатывают в разные моменты времени.
 Их дисперсия показана на рис. \ref{setup.tof-schem} (справа) и составляет $\sim1.8$\,нс.
 Поэтому время пролёта
 вычисляется как полусумма значений TOF\,L и R,
 что даёт временное разрешение $\sim300$\,пс.
 Информация о времени пролёта используется
 совместно с магнитным анализом
 (рис.~\ref{setup.tof2dimentions}).

 Система TOF необходима в наших исследованиях
 для исключения вклада неупругой реакции $np\to d+\pi^0$,
 в которой нейтрон захватывает протон, а мезон уносит избыточную энергию.
 Кинематика процесса такова, что в тех случаях,
 когда $\pi$-мезон в системе центра масс $np$-взаимодействия вылетает назад,
 импульс дейтрона оказывается немного больше импульса налетающего нейтрона,
 поэтому дейтроны попадают прямо под упругий пик протонов перезарядки $np\to pn$.
 Переход к мишени D$_2$ добавляет ещё один канал реакции $nn_d\to d+\pi^-$,
 где $n_d$ --- нейтрон ядра дейтерия. Поэтому примесь дейтронов возрастает вдвое.
 Например, при T$_n=550$\,МэВ  фон дейтронов достигает $\sim40$\,\%
 от пика протонов (рис.~\ref{setup.tof2dimentions}).

 \subsection{Детектор Окружения Мишени}\label{chapter-DTS}
\begin{figure}[!ht]
 \centering
 \scalebox{.44}{\includegraphics{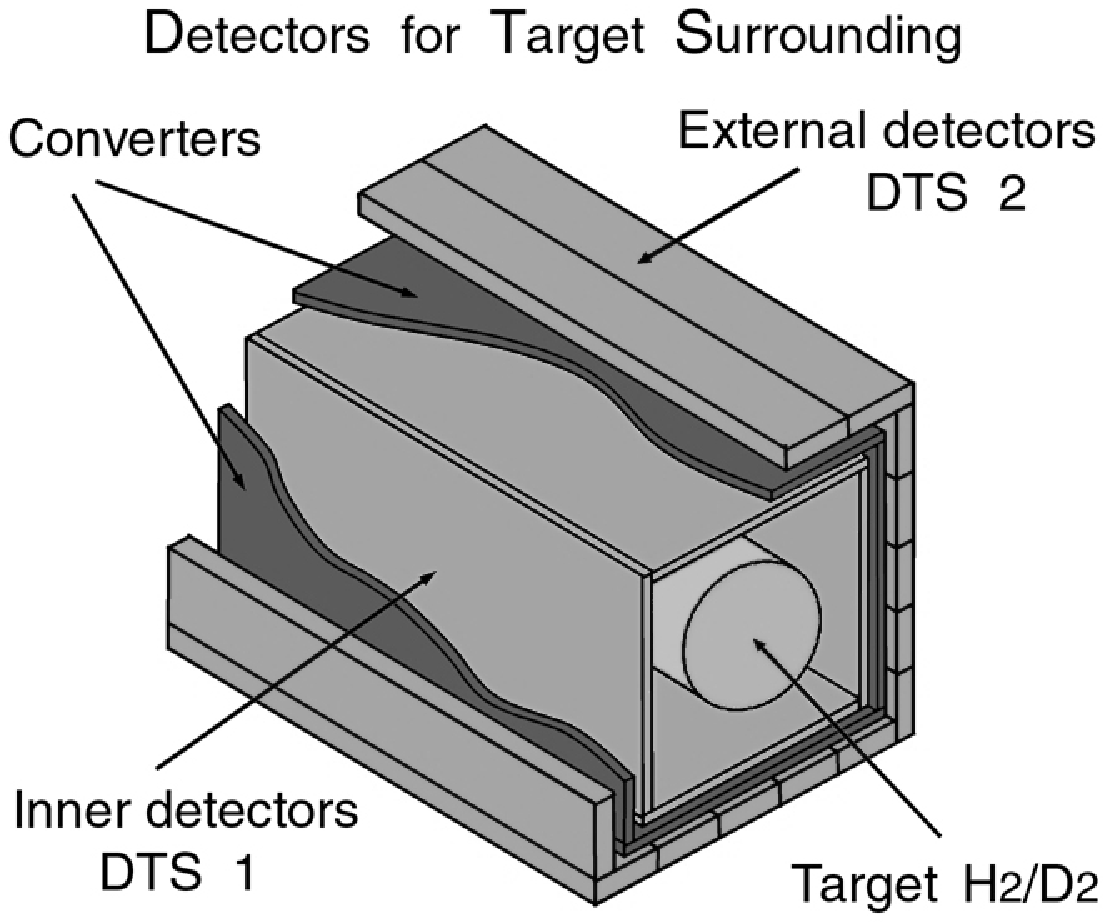}}
 \caption{\footnotesize Общий вид Детектора Окружения Мишени}\label{dts.5}
 \end{figure}

\begin{figure}[!ht]
 \centering
 \scalebox{.25}{\includegraphics{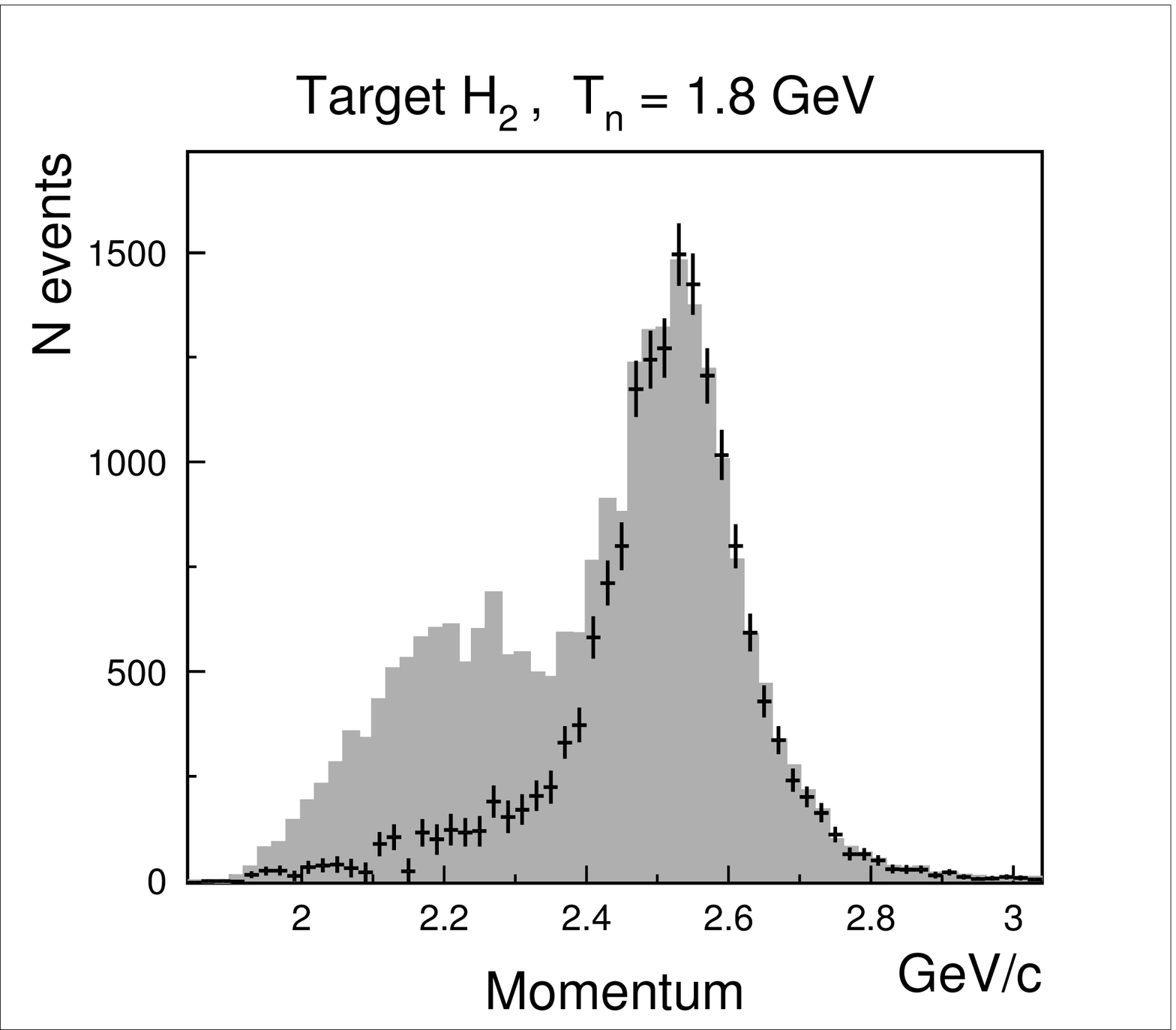}}\quad
 \scalebox{.25}{\includegraphics{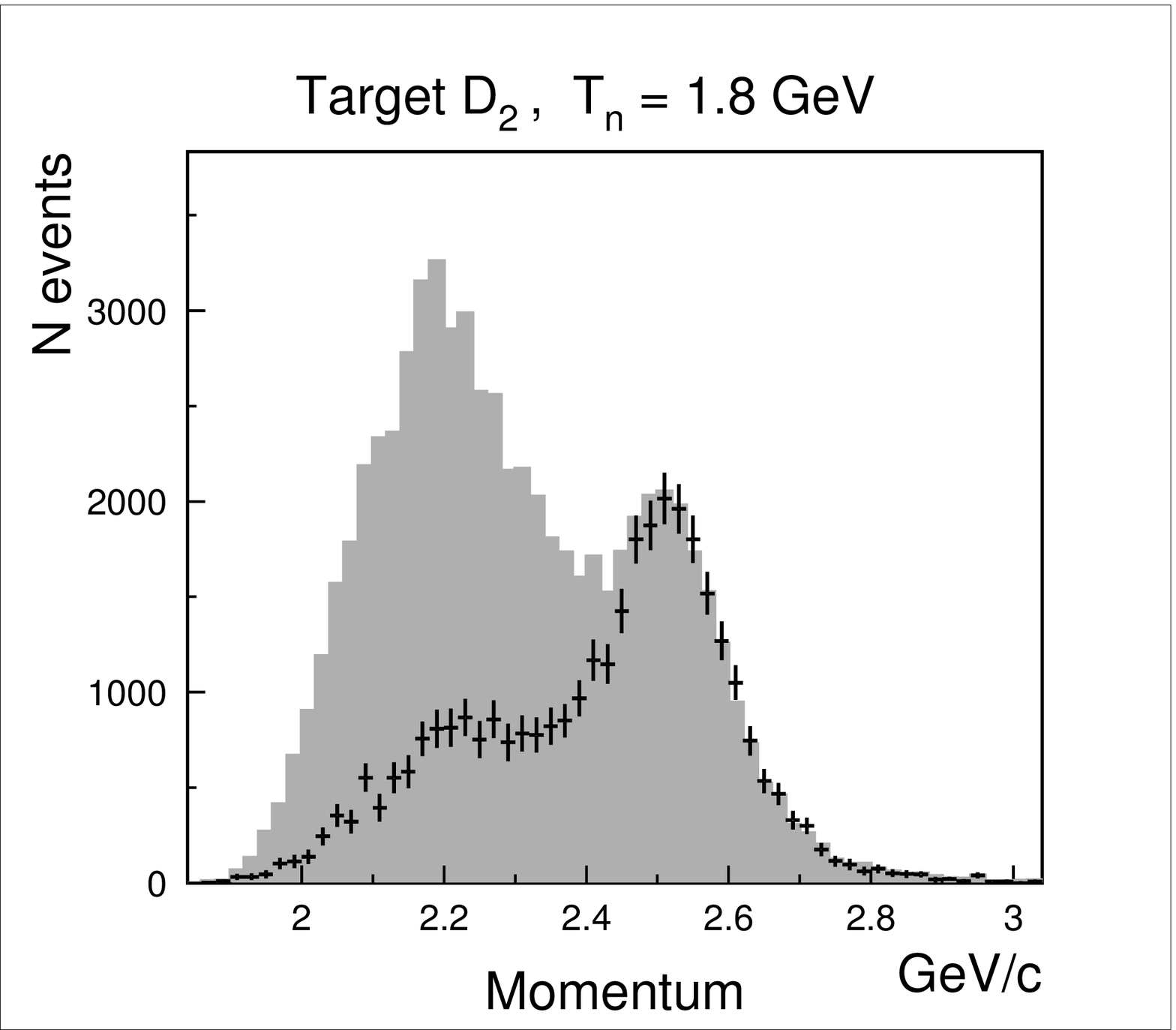}}
 \caption{\footnotesize Спектры импульсов протонов $n\to p$ перезарядки
 на H$_2$ и D$_2$ мишенях при энергии T$_n=1.8$\,ГэВ.
 Гистограммы сплошного серого цвета представляют спектры,
 полученные при участии только TOF системы.
 Гистограммы, изображённые с величинами ошибок, это те же спектры,
 но с учётом работы вето-системы ДОМ.
 }\label{dts.17}
 \end{figure}

 Препятствием для изучения спектров
 упругой $np\to pn$ и квазиупругой $nd\to p(nn)$ реакций перезарядки
 являются неупругие процессы, в диапазоне энергий T$_n=0.55\div2$\,ГэВ
 связанные в основном с возбуждением изобары $\Delta$\,(1232).
 Первоначально для разделения вкладов упругих и неупругих реакций
 использовался метод фита двойной функцией Гаусс + Брейт-Вигнер.
 Но при энергиях T$_n\geq1.4$\,ГэВ обнаружилась сильная зависимость
 оценки числа упругих событий под Гауссом
 от пределов наложения фита: $\Delta{N}\pm15$\,\%.
 Причина такого колебания заключается в том,
 что аппроксимация неупругой части спектра функцией Брейта-Вигнера
 не учитывает все каналы рождения $\Delta$-изобары
 и пригодна только для случая,
 когда резонанс происходит с мишенным нуклоном.
 Если же в $\Delta$ превращается налетающая частица,
 то протон, образующийся в процессе её распада,
 имеет более размытое распределение по импульсу
 и создаёт такой вклад в область неупругого пика,
 который нельзя аппроксимировать функцией Брейта-Вигнера.
 Решение проблемы было достигнуто аппаратным способом
 с помощью детектора окружения мишени ДОМ \cite{Shindin-DTS},
 который предназначен для регистрации протонов отдачи, $\pi$-мезонов
 и $\gamma$-квантов, которые появляются в распадах $\Delta$-резонансов.
 ДОМ состоит из трёх слоёв (рис.~\ref{dts.5}).
 \noindent
 Первый (внутренний) слой служит для регистрации
 заряженных $\pi$-мезонов и протонов
 и состоит из 4-х сцинцилляционных счётчиков,
 собранных по типу коробки без торцов,
 внутри которой размещаются криогенные H$_2$/D$_2$
 или твёрдые CD$_2$/CH$_2$/C-мишени.
 Второй слой из металлических пластин
 необходим для конверсии $\gamma$-квантов: $\gamma\to e^+ + e^-$.
 Толщина пластин составляет оптимум в 1.5 радиационные единицы.
 Третий (внешний) слой состоит из 20-ти сцинцилляционных счётчиков,
 которые собраны по 5 на каждую сторону и
 предназначены для регистрации вторичных электронов
 от конверсии $\gamma$-квантов, а также для
 $\pi^+$ и $\pi^-$ мезонов, имеющих достаточную энергию
 для прохождения сквозь конвертеры.
 В наших расчётах способность регистрации
 заряженных частиц получила оценку 92\,\%, а нейтральных --- 67\,\%.
 Это даёт суммарную эффективность $\sim80$\,\%,
 что и подтвердилось опытным путём:
 спектры, полученные в антисовпадении с сигналом детектора ДОМ,
 показали (рис.~\ref{dts.17}) подавление неупругого фона с фактором $\approx5$.

 \section{Смещение квазиупругого пика}
\begin{figure}[!ht]
  \centering
  \scalebox{.25}{\includegraphics{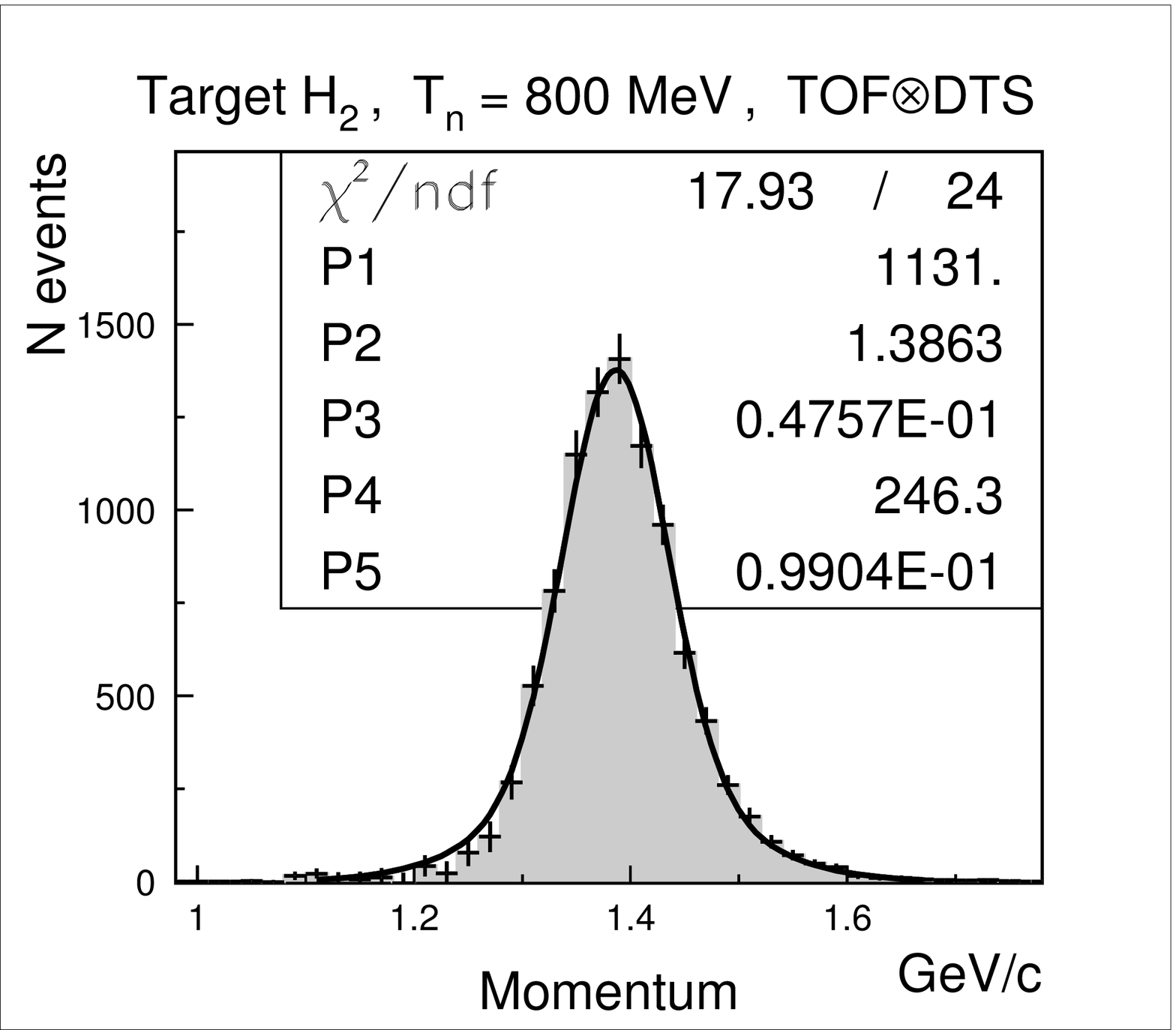}}\quad
  \scalebox{.25}{\includegraphics{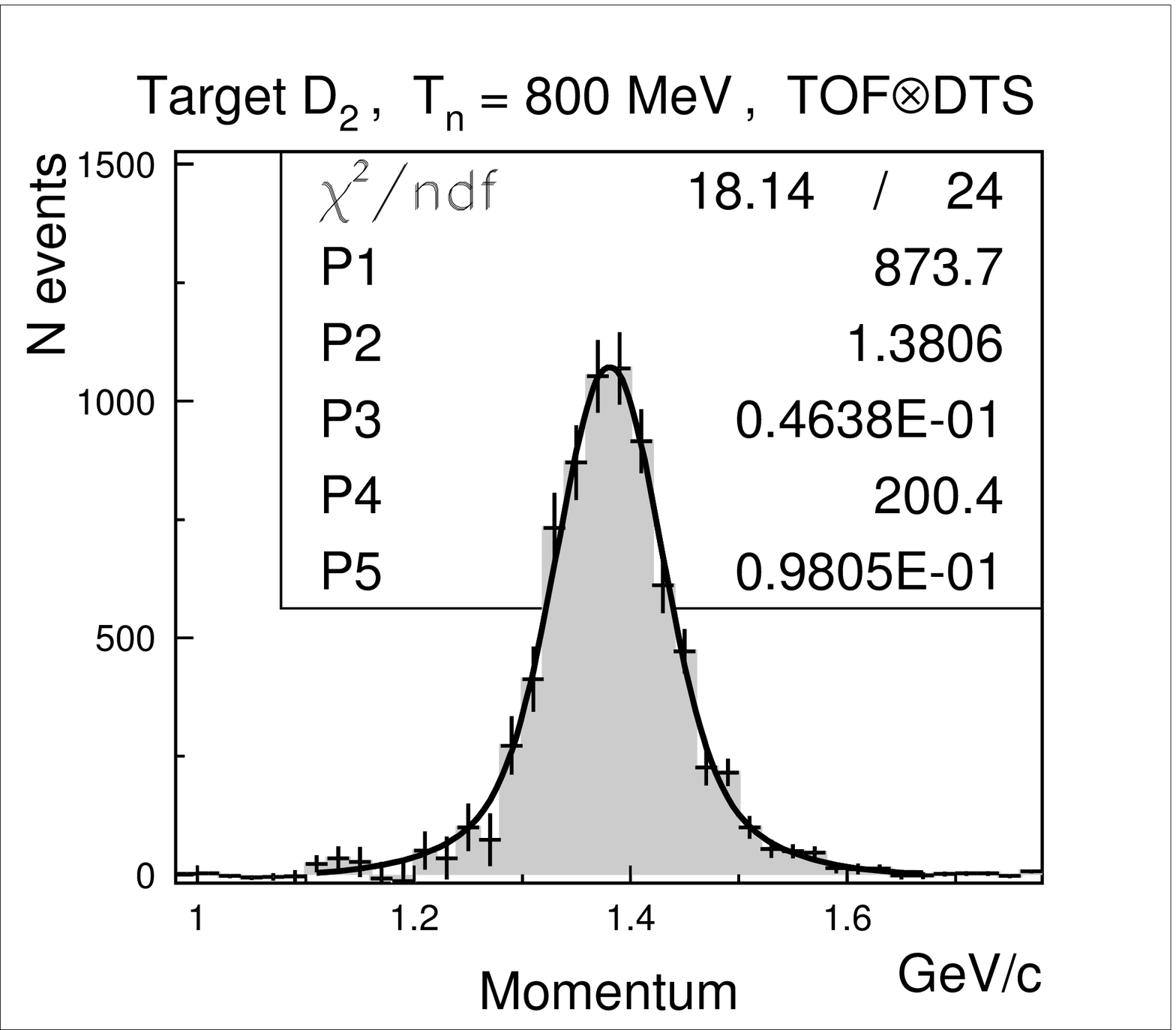}}
  \caption{\footnotesize {Слева спектр импульсов протонов
  $np\to pn$ перезарядки под нулём при энергии T$_n$\,=\,0.8\,ГэВ.
  Справа представлен спектр квазиупругой реакции $nd\to p(nn)$.
  Оба спектра аппроксимируются функцией \eqref{setup.2gauss-function}.
  Параметры фита:
  p1 -- высота первого Гаусса;
  p2 -- позиция центра;
  p3 -- ср. кв. отклонение первого Гаусса;
  p4 -- высота второго Гаусса
  и p5 -- его ср.\,кв. отклонение.
  Величина сдвига квазиупругого пика в сторону меньших значений составляет
  $\delta P=\textrm{p2}(\textrm{H}_2)-\textrm{p2}(\textrm{D}_2)=5.7\pm2.4$\,МэВ/$c$.
  }}\label{shifting.respectively shift at 0.8 GeV}
  \end{figure}

  Наиболее ярко особенность квазиупругой реакции
  проявляется при энергиях, где фон неупругих событий,
  связанных с возбуждением резонанса $\Delta$\,(1232), пренебрежимо мал.
  Например, при T$_n=0.8$\,ГэВ спектры практически свободны от этих вкладов
  и представляют чистые процессы $np\to pn$ и $nd\to p(nn)$.
  Распределения имеют одну и ту же форму
  (рис.~\ref{shifting.respectively shift at 0.8 GeV}),
  \noindent
  но спектр протонов перезарядки на дейтерии
  смещён почти на 6\,МэВ/$c$ в сторону меньших значений.
  Вещество мишени менялось каждые 3 часа по схеме: CH$_2\to$ CD$_2\to$ C,
  то есть набор шёл при одних и тех же условиях
  и систематическая ошибка здесь сведена к минимуму.
  Ширина бина в гистограммах выбрана в 20\,МэВ/$c$ при том,
  что разрешение по импульсу при энергии T$_n$\,=\,0.8\,ГэВ не хуже 10\,МэВ/$c$.
  Для аппроксимации спектров упругой и квазиупругой реакций используется функция
  двойного нормального распределения:
  \begin{equation}\label{setup.2gauss-function}
   f(p) \;=\; C_1\exp\left(\frac{-(p-M)^2}{2\,\sigma^2_1}\right)\,+\;
   C_2\exp\left(\frac{-(p-M)^2}{2\,\sigma^2_2}\right)\;.
  \end{equation}
  Оба Гаусса \eqref{setup.2gauss-function} имеют одинаковые мат. ожидания.
  Фит каждого спектра определяет 5 свободных параметров c $\chi^2\approx1$
  (рис.~\ref{shifting.respectively shift at 0.8 GeV} или
  прил.~\ref{Appendix.neutron momentum distribution},
  рис.~\ref{appendix.neutron momentum distribution at 1.0 GeV}).
  Смещение между пиками упругой $np\to pn$ и квазиупругой $nd\to p(nn)$ реакций перезарядки
  находится как разность положений их центров
  $M_{\textrm{H}_2}$ и $M_{\textrm{D}_2}$ соответственно:
  $\delta{P}=M_{\textrm{H}_2}-M_{\textrm{D}_2}$.

\begin{figure}[!ht]
  \centering
  \scalebox{.25}{\includegraphics{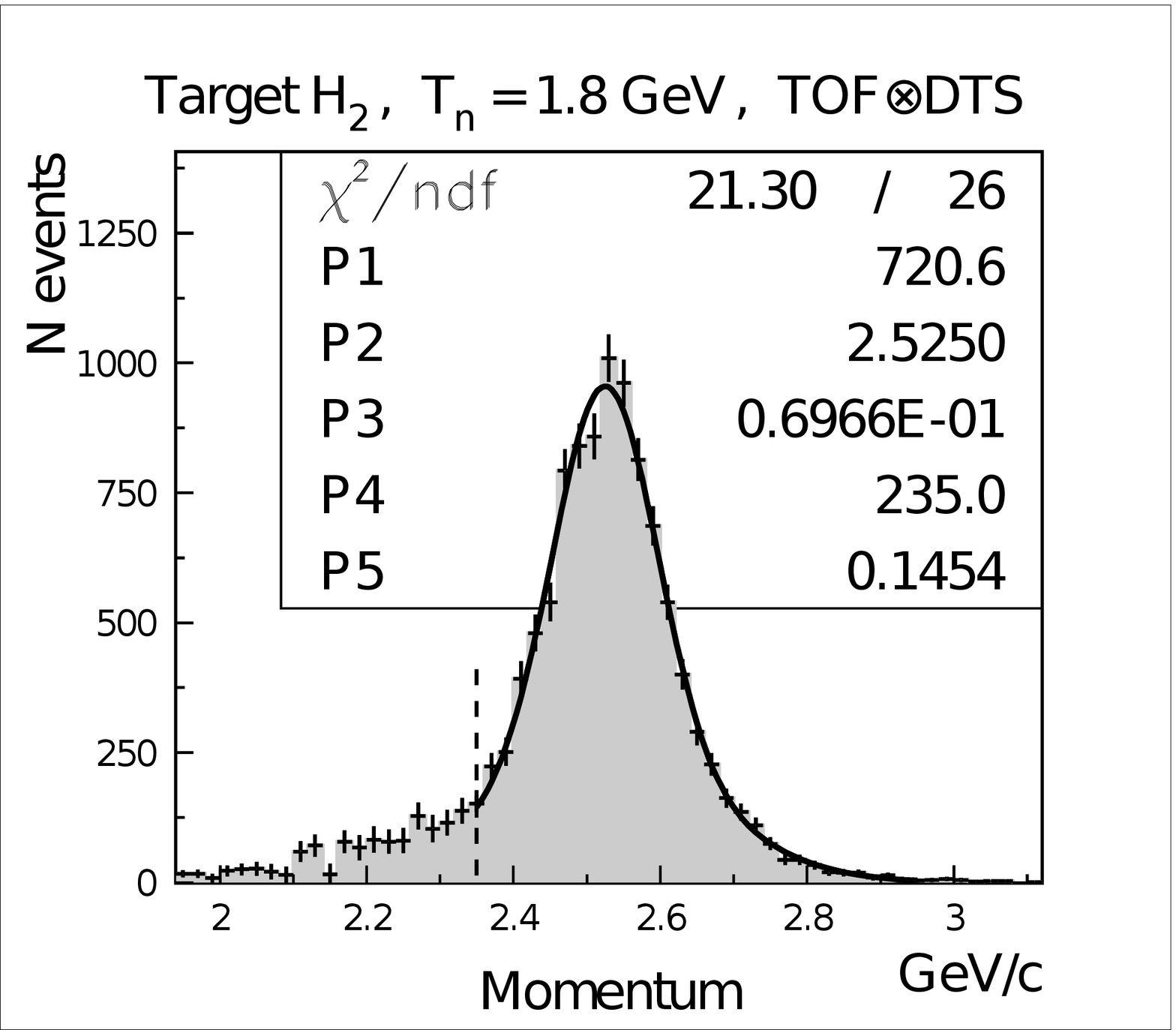}}\quad
  \scalebox{.25}{\includegraphics{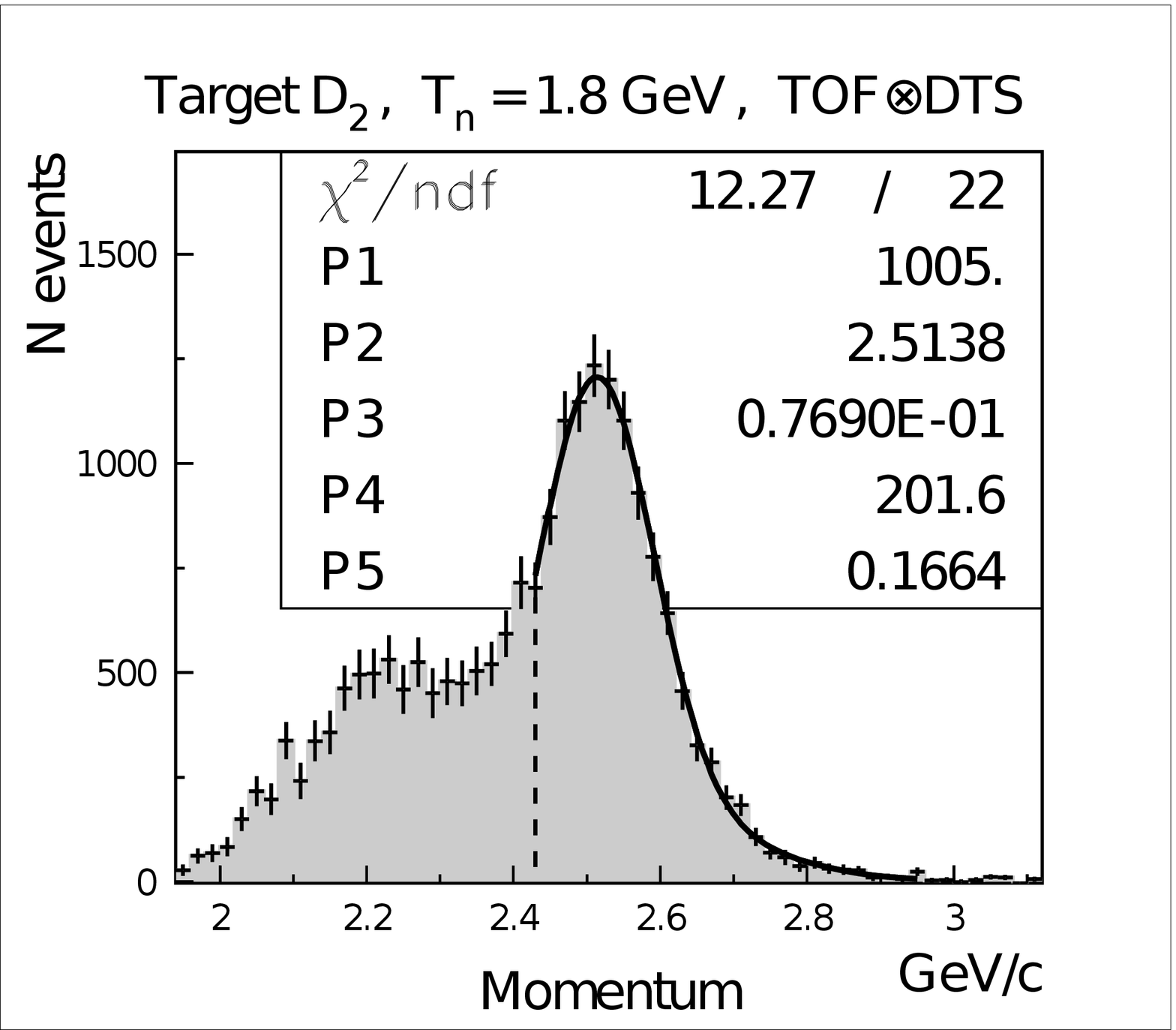}}
  \caption{\footnotesize {То же самое, что и на рис. \ref{shifting.respectively shift at 0.8 GeV},
  но при большей энергии T$_n$\,=\,1.8\,ГэВ.
  Вертикальные пунктирные линии на обеих гистограммах
  показывают начало фита функцией \eqref{setup.2gauss-function}.
  Сдвиг между пиками равен
  $\delta P=\textrm{p2}(\textrm{H}_2)-\textrm{p2}(\textrm{D}_2)=11\pm3.4$\,МэВ/$c$.
  }}
  \label{shifting.respectively shift at 1.8 GeV}
  \end{figure}

\begin{figure}[!ht]
  \centering
  \scalebox{.25}{\includegraphics{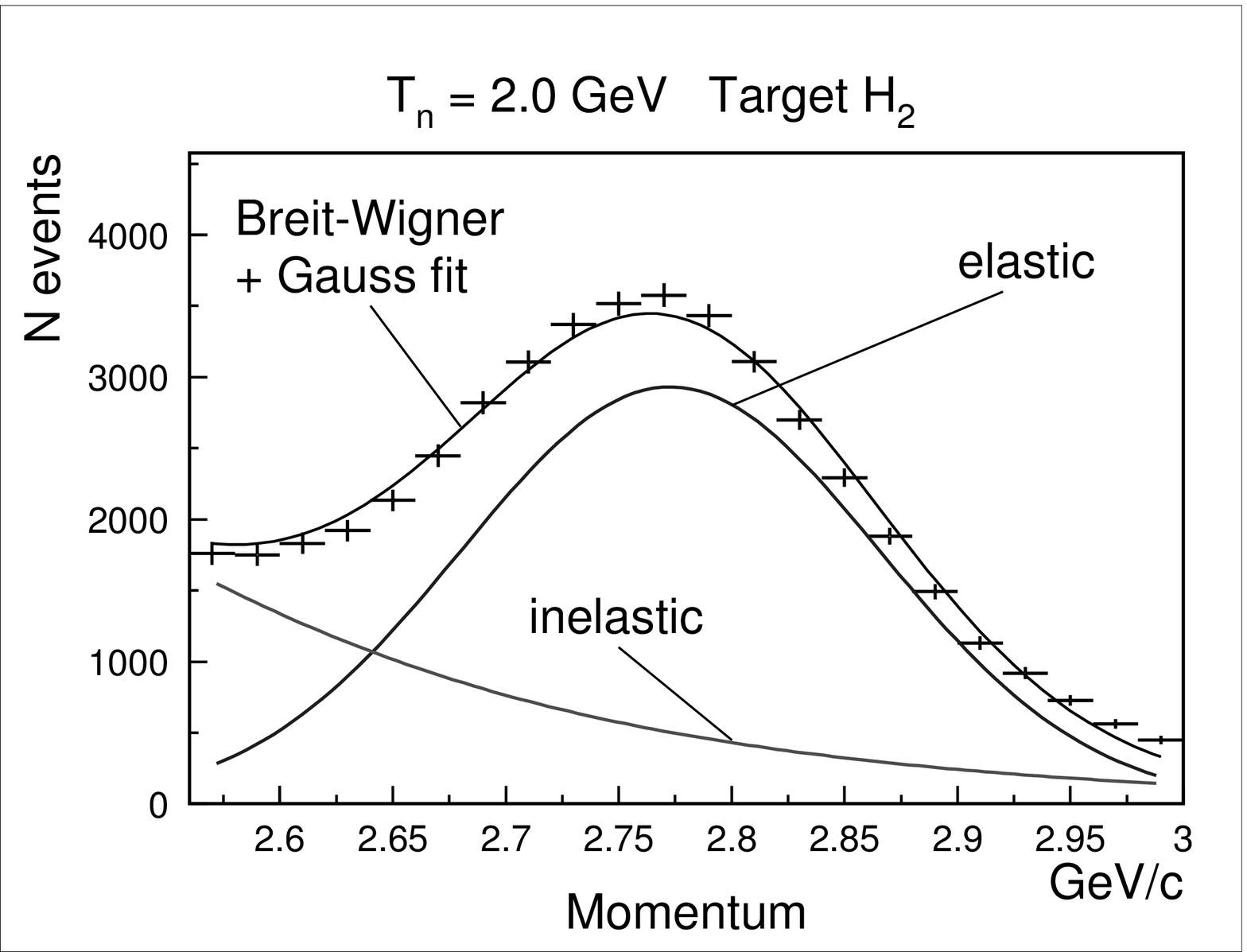}}
  \caption{\footnotesize {Спектр протонов $n\to p$-перезарядки при энергии T$_n=2.0$\,ГэВ,
  полученный без учёта сигналов вето-системы ДОМ.
  Результат аппроксимации двойной функцией Брейт-Вигнер + Гаусс
  разделен на два вклада inelastic и elastic.
  Неупругие события образуют склон,
  что приводит к фиктивному смещению вершины пика
  упругой перезарядки на величину $\sim5$\,МэВ/$c$ .}}
  \label{shifting.shift by inelastic influence}
  \end{figure}

  При более высоких энергиях 
  фон неупругих событий возрастает.
  Для примера на рис. \ref{shifting.respectively shift at 1.8 GeV}
  приведены спектры протонов при T$_n$\,=\,1.8\,ГэВ.
  \noindent
  Чтобы уменьшить влияние фоновых событий,
  которые остались в спектре протонов от H$_2$-мишени
  несмотря на их 5-ти кратное подавление вето-системой ДОМ,
  фит функцией \eqref{setup.2gauss-function} ограничен слева.
  На D$_2$-мишени ситуация усложняется,
  поскольку вклад неупругих событий увеличивается, а выход "<упругих"> падает:
  баланс между упругими и неупругими изменяется почти в 4 раза
  по сравнению с мишенью H$_2$.
  Сдвиг между спектрами определяется на уровне $\delta P$\,=\,11\,МэВ/$c$.
  Однако в значении $\delta P$ появляется фиктивная часть: события неупругих реакций
  образуют склон под пиком и его наивысшая точка смещается влево
  (рис.~\ref{shifting.shift by inelastic influence}),
  причём, для пика протонов реакции $nd\to p(nn)$
  \noindent
  такого рода фактор действует заведомо сильнее.
  Очевидно, что при энергиях T$_n\geq1.4$\,ГэВ
  мы снова возвращаемся к вопросу описания фона неупругих реакций,
  связанных с рождением и распадом резонанса $\Delta$\,(1232).
  Поэтому требовался другой способ, позволяющий нивелировать эту проблему.

 \section{Метод параметризации сдвига} 
 Возникла идея: искать относительный сдвиг $\delta{P}$,
 не прибегая к аппроксимации спектров упругой $np\to pn$
 и квазиупругой $nd\to p(nn)$ реакций перезарядки.
 Так как спектры имеют одинаковую\footnote{Данные
 при T$_n$\,=\,0.55 и 0.8 ГэВ,
 где фон неупругих событий пренебрежимо мал,
 показали, что упругий и квазиупругий спектры изоморфны,
 хотя ширина пика на дейтерии могла бы оказаться больше
 за счёт размывки по импульсам Ферми.} форму
 (рис.~\ref{shifting.respectively shift at 0.8 GeV}),
 если их разделить друг на друга,
 за счёт сдвига $\delta{P}$ на результирующей гистограмме появится наклон
  \begin{figure}[!ht]
  \centering
  \scalebox{.25}{\includegraphics{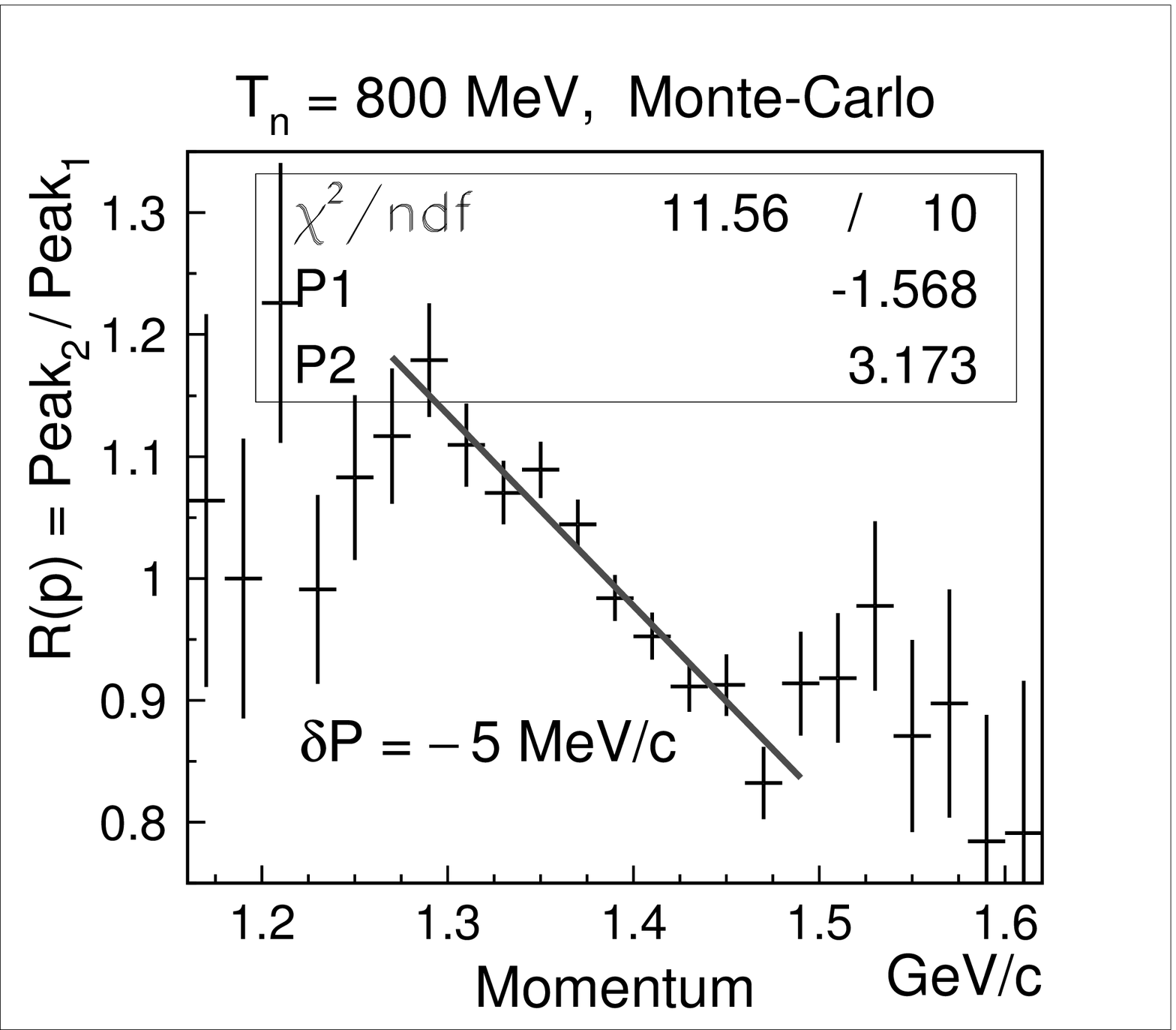}}\quad
  \scalebox{.25}{\includegraphics{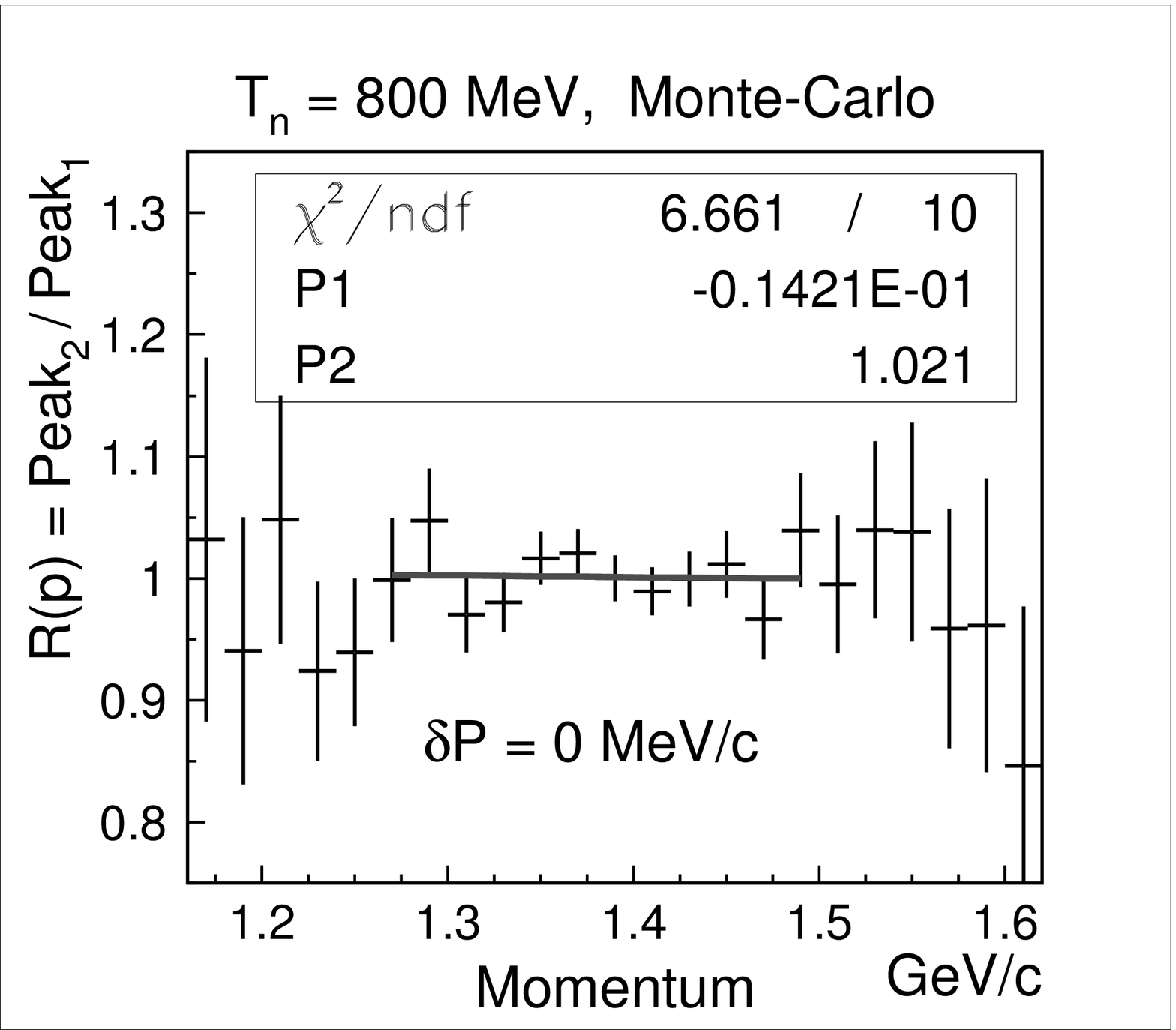}}
  \caption{\footnotesize {Гистограммы показывают результаты деления
  одного пика на другой при двух разных сдвигах второго пика относительно первого.
  Левый рисунок получен при отрицательном сдвиге $\delta{P}=-5$\,МэВ/$c$,
  правый --- при нулевом.
  Исходные спектры построены в рамках Монте-Карло модели.
  Они имеют одинаковую форму двойного нормального распределения
  \eqref{setup.2gauss-function}.
  Для определения наклона используется область $\sim200$\,МэВ/с,
  которая фитируется прямой линией $y=\textrm{p}_1\cdot x+\textrm{p}_2$.
  }}\label{shifting.shifts between two peaks}
  \end{figure}\noindent
 (рис.~\ref{shifting.shifts between two peaks}, слева). 

\begin{figure}[!ht]
  \centering
  \scalebox{.25}{\includegraphics{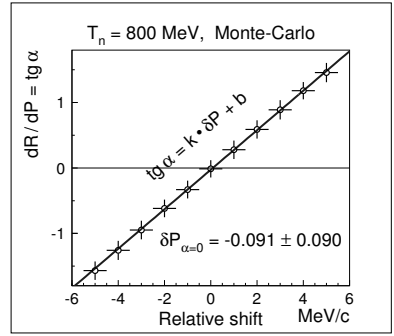}}
  \caption{\footnotesize {Показан принцип
  определения сдвига между пиками одинаковой формы.
  Данные нескольких итераций, две из которых изображены
  на рис. \ref{shifting.shifts between two peaks},
  представлены так:
  по оси абсцисс отложены величины сдвигов $\delta{P}$,
  а по оси ординат значения тангенса угла наклона $\tg{\alpha}$.
  Распределение фитируется прямой $y=kx+b$,
  после чего определяется точка, где параметр $\tg{\alpha}$ принимает нулевое значение,
  что соответствует положению, когда центры пиков совпадают.
  }}\label{shifting.principle of optimal shift definition}
  \end{figure}

 \noindent
 Причину этого легко понять из расчётов:
 \begin{eqnarray}\label{shifting.relation of Gauss-D2 / Gauss-H2}
   R(p) = \dfrac{\textrm{Гаусс}_{\,\textrm{D}_2}}{\textrm{Гаусс}_{\,\textrm{H}_2}} =\,
   \exp{\dfrac{-(p-M_{\textrm{H}_2})\,\delta{P}}{2\,\sigma^2}}\;,
 \end{eqnarray}
 где: $\delta{P}=M_{\textrm{H}_2}-M_{\textrm{D}_2}$
 и $\sigma=\sigma_{\textrm{D}_2}=\sigma_{\textrm{H}_2}$.
 В тех случаях, когда $\delta{P}\ll\sigma$,
 а на практике так и есть,
 поскольку средне-квадратичное отклонение десятикратно превосходит сдвиг,
 распределение $R(p)$ вблизи $M_{\textrm{H}_2}$ будет подобно прямой линии,
 наклон которой пропорционален сдвигу:
 \begin{eqnarray}\label{shifting.tangent of Gauss-H2 / Gauss-D2}
   \tg{\alpha}\;=\; \frac{dR}{dp}
   \;\simeq\;-\delta{P}\cdot\frac{1}{2\,\sigma^2}\;.
 \end{eqnarray}
 Для пиков любой гладкой формы
 при их малых относительных смещениях действует то же правило,
 и если сдвига нет, наклон в их отношении будет нулевым.
 Величину $\sigma=\sigma_{\textrm{H}_2}$ определять не обязательно.
 Достаточно сместить один из спектров и снова провести деление.
 По значениям $\tg\alpha$ делается экстраполяция в точку,
 где производная $\,{dR}/{dp}\,$ обращается в ноль.
 Метод использует 10 перемещений с шагом 1\,МэВ/$c$,
 и уже на основе всех измерений ищется значение $\delta{P}_{\alpha=0}$
 (рис.~\ref{shifting.principle of optimal shift definition}).

  \subsection{Определение D$_2$/H$_2$-отношения по правым \\ половинам упругого и квазиупругого пиков}
\begin{figure}[!ht]
  \centering
  \scalebox{.25}{\includegraphics{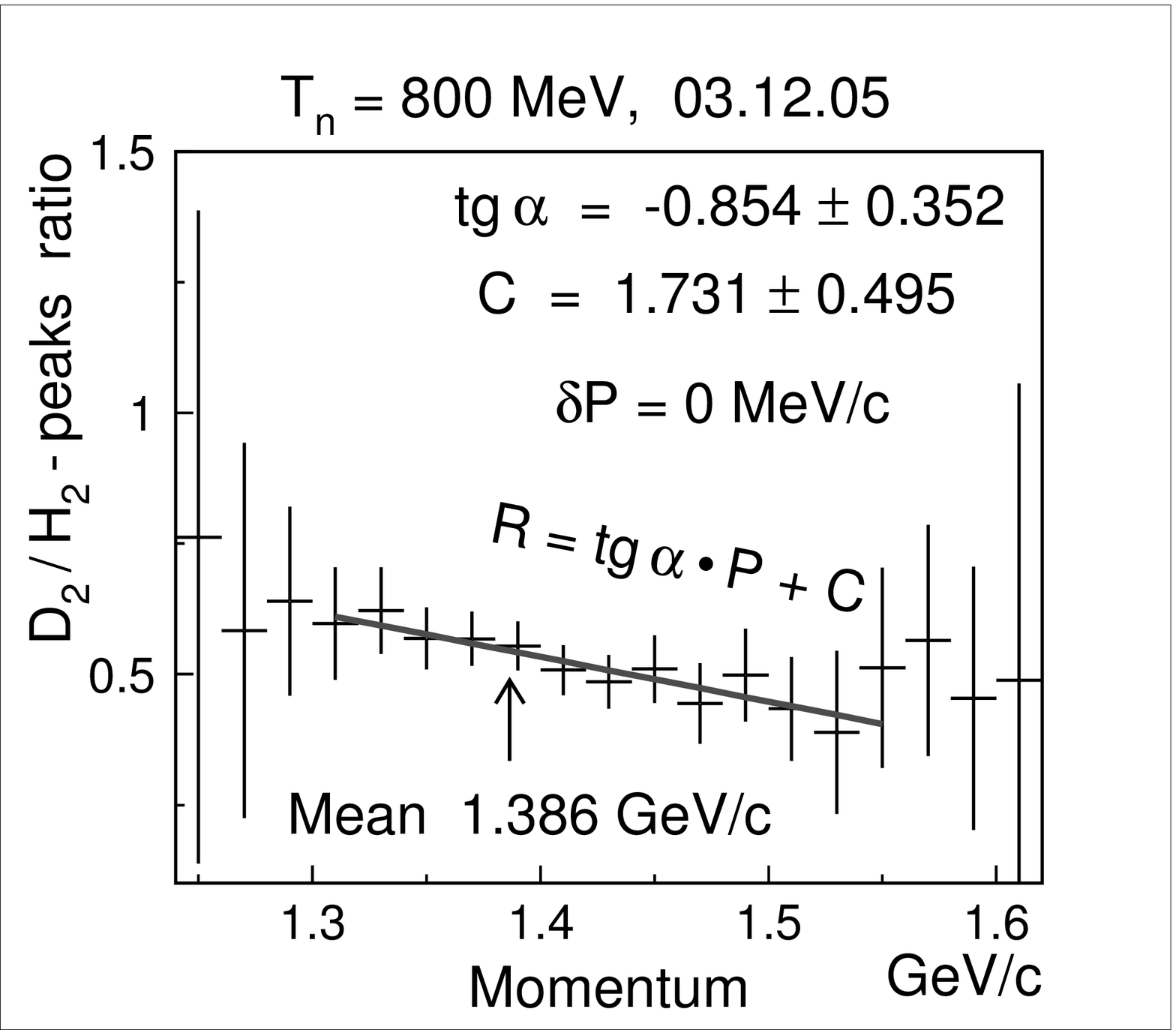}}\quad
  \scalebox{.25}{\includegraphics{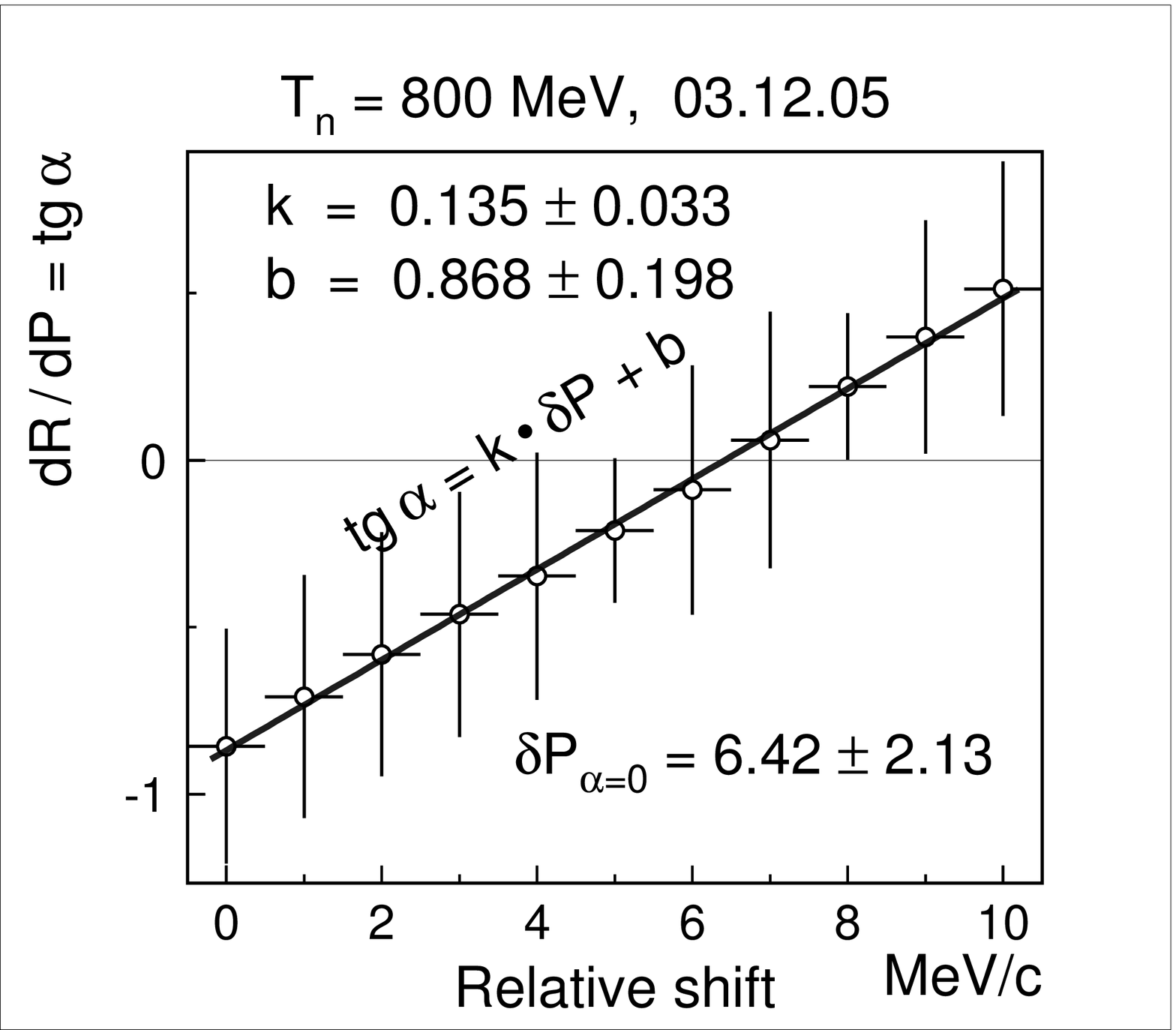}}
  \caption{\footnotesize {Гистограмма слева представляет отношения выходов
  протонов в реакциях $n\to p$ перезарядки на водороде и дейтерии при рассеянии протонов под нулём градусов.
  Наблюдаемый склон фитируется прямой линией $R=\tg{\alpha}\cdot{P}+C$
  в пределах от $-\,1.5\sigma_1$ до $+\,3\sigma_1$ относительно центра упругого пика,
  положение которого указывает стрелка.
  Здесь $\sigma_1\sim50$\,МэВ/$c$ --- ср. кв. отклонение основного Гаусса
  (параметр p3 на рис.~\ref{shifting.respectively shift at 0.8 GeV},
  \ref{shifting.respectively shift at 1.8 GeV}).
  Вторая гистограмма показывает
  линейную зависимость $\tg{\alpha}$ от переменной $\delta{P}$,
  которая изменяется с шагом 1\,МэВ/$c$.
  Параметр $\tg{\alpha}$ обращается с ноль
  в точке $\delta{P}_{\alpha=0}=6.42$\,МэВ/$c$,
  что определяет сдвиг между спектрами упругой $np\to pn$
  и квазиупругой $nd\to p(nn)$ реакций.
  }}\label{shifting.shifting at 800 MeV/c}
  \end{figure}

\begin{figure}[!ht]
  \centering
  \scalebox{.25}{\includegraphics{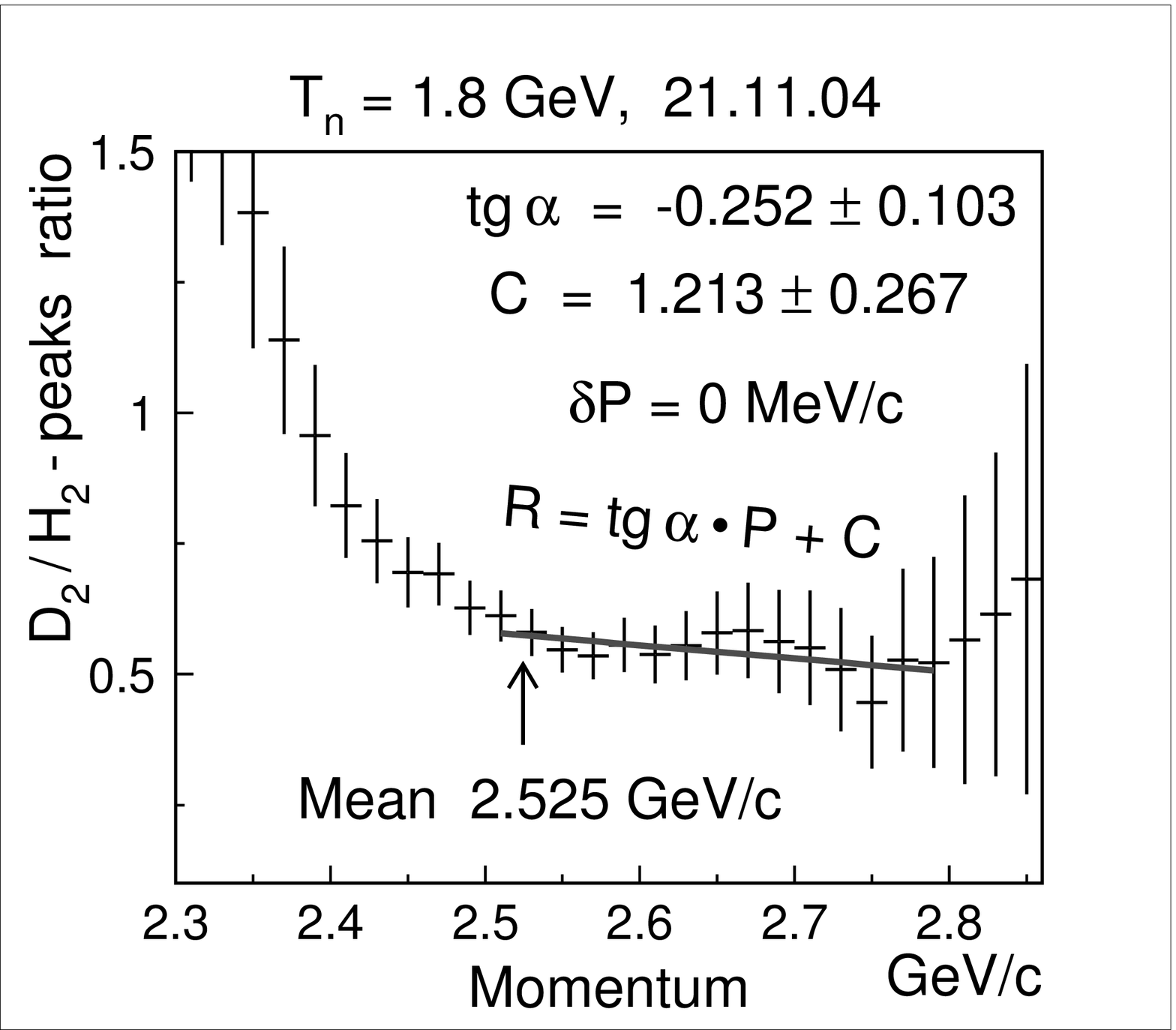}}\quad
  \scalebox{.25}{\includegraphics{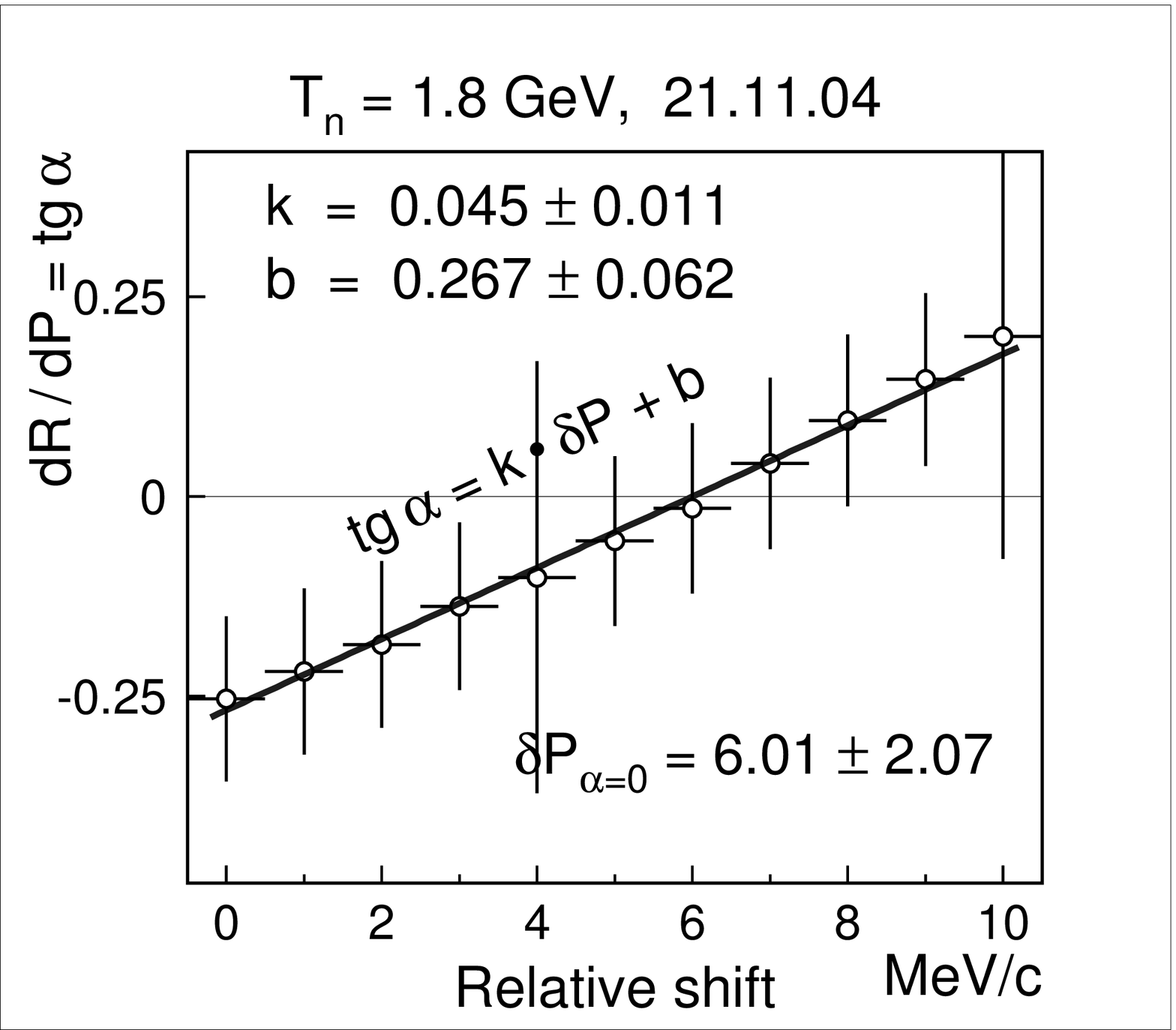}}
  \caption{\footnotesize {Гистограммы представляют то же самое,
  что и на рис.\,\ref{shifting.shifting at 800 MeV/c},
  но при энергии T$_n$\,=\,1.8\,ГэВ.
  В D$_2$/H$_2$-отношении в левую сторону
  от центрального значения Mean\,$\approx$\,2.53\,ГэВ/$c$ наблюдается подъём,
  который связан почти с 2-х увеличением
  числа неупругих событий на дейтерии
  относительно числа неупругих событий на водороде.
  Параметр наклона $\tg{\alpha}$ определяется
  в границах плато от $\textrm{Mean}-0.5\sigma_1$ до $\textrm{Mean}+3\sigma_1$.
  Координата точки, в которой $\tg{\alpha}$ обращается в ноль,
  соответствует значению сдвига $\delta{P}_{\alpha=0}=6.01$\,МэВ/$c$.
  }}\label{shifting.shifting at 1.8 GeV/c}
  \end{figure}

  Метод параметризации сдвига был проверен при энергиях T$_n=550$ и 800\,МэВ,
  где фон неупругих реакций пренебрежимо мал и где выполняется
  основное условие подобия пиков.
  На рис. \ref{shifting.shifting at 800 MeV/c}
  показаны результаты метода при энергии T$_n=800$\,МэВ.
  \noindent
  Значение $\delta{P}$ определяется на уровне $6.4\pm2.1$\,МэВ/$c$,
  что согласуется с результатом аппроксимации спектров функцией
  \eqref{setup.2gauss-function}.
  При переходе к б\'{о}льшим энергиям фон неупругих
  реакций возрастает, причём на дейтерии
  выход неупругих увеличивается почти вдвое,
  поэтому левые части спектров
  (рис.~\ref{shifting.respectively shift at 1.8 GeV})
  становятся существенно разными.
  В D$_2$/H$_2$-отношении\footnote{
  Для приведения к общему знаменателю опытов с D$_2$ и H$_2$-мишенями
  в момент деления квазиупругого и упругого спектров друг на друга
  каждый нормируется на эффективность установки и берётся со своей поправкой
  1/M$_{\textrm{D}_2}$ и 1/M$_{\textrm{H}_2}$ соответственно,
  где: M$_{\textrm{D}_2}$ и M$_{\textrm{H}_2}$ --- показания мониторных счётчиков
  (рис.~\ref{setup.spectrometer}).
  Также учитывается разница в количестве ядер водорода и дейтерия.}
  (рис.~\ref{shifting.shifting at 1.8 GeV/c})
  \noindent
  слева от центрального значения $\textrm{Mean}=P_n$ импульса пучка нейтронов
  возникает переходная область, где наблюдается резкий подъём,
  который соответствует изменению баланса между
  вкладами упругих и неупругих реакций $n\to p$ перезарядки на дейтерии и водороде.
  Метод параметризации применим здесь только для {\it хорошей}
  части D$_2$/H$_2$-отношения, где число неупругих событий падает
  вследствие кинематического предела рождения изобары $\Delta$\,(1232),
  то есть правее значения $P_n$.
  Тем не менее, спектр нейтронов имеет достаточно протяжённые {\it хвосты}
  по обе стороны от  $P_n$ (прил.~\ref{Appendix.neutron momentum distribution},
  рис.~\ref{appendix.neutron momentum distribution at 1.0 GeV}),
  и некоторая доля неупругих всё равно проходит в эту {\it хорошую} половину.
  Для определения их числа были проведены расчёты
  в рамках Монте-Карло модели спектрометра
  (таб.~\ref{shifting.inelastic yields on the right sides of H2 & D2 elastic peaks.table}).
  \begin{table}[!ht]
  \caption{\small Процентная доля неупругих событий в правой половине спектра
  протонов в реакциях перезарядки $np\to pn$ и $nd\to p(nn)$.}
  \label{shifting.inelastic yields on the right sides of H2 & D2 elastic peaks.table}
  \centering
  \begin{tabular}{|c|c|c|c|c|c|c|c|c|}
  \hline
  T$_n$, ГэВ  &  0.55 & \;0.8\; & \;1.0\; & \;1.2\; & \;1.4\; & \;1.7\; & \;1.8\; & \;2.0\; \\ \hline
  H$_2$ & 0.23 & 0.04 & 0.18 & 0.38 & 0.72 & 1.82 & 2.30 & 2.73   \\ \hline
  D$_2$ & 0.65 & 0.11 & 0.66 & 1.62 & 2.70 & 7.19 & 8.33 & 10.33  \\ \hline
  \multicolumn{9}{c}{\small С учётом работы Детектора Окружения Мишени.}  \\ \hline
  H$_2$ & 0.05 & 0.01 & $-$  & $-$  & 0.20 & 0.49 & 0.68 & 0.85   \\ \hline
  D$_2$ & 0.16 & 0.03 & $-$  & $-$  & 0.79 & 2.54 & 2.69 & 3.64   \\ \hline
  \end{tabular}
  \end{table}

  Оказалось, что в диапазоне энергий T$_n=0.55\div2.0$\,ГэВ
  при условии работы вето-системы ДОМ вкладами неупругих реакций можно пренебречь
  и определять сдвиг $\delta{P}$ по {\it хорошей} части D$_2$/H$_2$-гистограммы.
  Так как ошибка сдвига составляет $\sim30$\,\% от величины $\delta{P}$
  (рис.~\ref{shifting.shifting at 800 MeV/c},~\ref{shifting.shifting at 1.8 GeV/c}),
  а влияние событий неупругих реакций даже при T$_n=2.0$\,ГэВ не превышает $3$\,\%,
  то поправка на этот фактор не существенна.

  Возможность находить смещение между пиками,
  используя лишь их половины, является преимуществом метода параметризации.
 \enlargethispage{1\baselineskip}
  Здесь не нужно ни угадывать форму распределения, ни заводить функцию фита за вершину спектра,
  чтобы оценить положение его центра.

  \subsection{Погрешность метода}
  В гистограмме D$_2$/H$_2$-отношения область фита имеет ширину более 200\,МэВ/$c$,
  а смещение $\delta{P}$ составляет от 0 до 10\,МэВ/$c$,
  поэтому в каждой итерации участвуют практически те же события.
  Это говорит о том, что все значения $\{\tg{\alpha}_n\}$ взаимосвязаны.
  Расхождение в оценках ошибок этого параметра
  (рис.~\ref{shifting.shifting at 800 MeV/c} и~\ref{shifting.shifting at 1.8 GeV/c})
  возникает за счёт разбиения событий по бинам гистограммы D$_2$-спектра,
  когда в последовательности итераций с шагом 1\,МэВ/$c$
  они перекладываются из одного бина гистограммы в другой.
  В предельном случае, если статистика неограничена,
  корреляция будет близка к 100\,\%,
  поэтому ошибки параметра наклона нужно суммировать арифметически.
  Получаем формулу:
  \begin{equation}\label{shifting.true shift dispersion}
    \sigma({\delta{P}}) =
    \frac{1}{k}\cdot\frac{1}{11}\sum\limits_{n=1}^{11}\sigma(\tg{\alpha}_n)\;.
  \end{equation}
  Мы полагаем, что величина $k$ имеет здесь точное значение,
  и возвращаемся к исходному источнику погрешности:
  средняя ошибка параметра $\tg{\alpha}$ отображается
  на горизонтальную ось сдвига $\delta{P}$.

  \section{Экспериментальные данные}
\begin{figure}[!ht]
  \centering
  \scalebox{.33}{\includegraphics{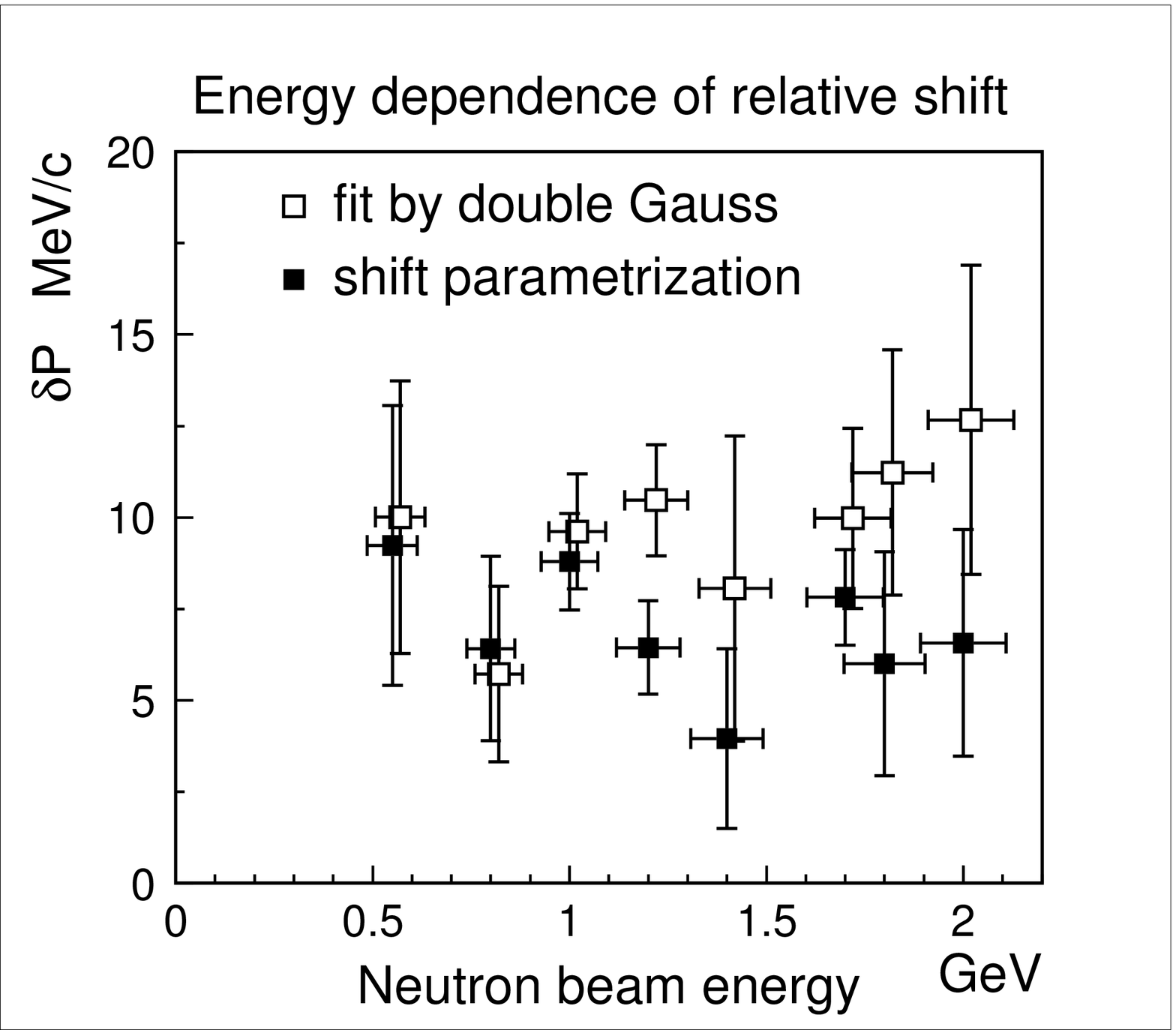}}
  \caption{\footnotesize {Энергетическая зависимость относительного смещения между
  упругим и квазиупругим спектрами протонов,
  полученных в реакциях перезарядки нейтрона на водороде и дейтерии.
  Представлены значения сдвигов, определённые двумя разными способами.
  Метод параметризации
  использует свободные от неупругих событий правые части спектров
  и даёт более кучное распределение вблизи значения $\delta{P}\sim6.5$\,МэВ/$c$.
  Другой способ проводит аппроксимацию
  упругого и квазиупругого пиков функцией \eqref{setup.2gauss-function}.
  Возрастание числа неупругих событий
  смещает вершину пика в левую сторону (рис.~\ref{shifting.shift by inelastic influence}),
  и в большей степени это касается квазиупругого пика.
  Поэтому такой метод даёт более высокие значения относительных сдвигов.
  }}\label{shifting.shifting experimental data}
  \end{figure}

  По результатам измерения реакции перезарядки нейтрона на водородной и дейтериевой мишенях
  в диапазоне энергий T$_n=0.55\div2.0$\,ГэВ при рассеянии
  вторичных протонов под нулём градусов
  и проведенного анализа их импульсных спектров оказалось,
  что смещение квазиупругого пика в сторону меньших значений
  от центра упругого составляет порядка 6.5\,МэВ/$c$
  (таб.~\ref{shifting.relative shifts and their errors.table}
  и рис.~\ref{shifting.shifting experimental data}).
  \begin{table}[!ht]
  \caption{\small Относительные сдвиги между пиками вторичных протонов
   упругой $np\to pn$ и квазиупругой $nd\to p(nn)$ реакций.
   Значения $\delta{P}_{2G}$ получены с помощью фита
   функцией \eqref{setup.2gauss-function},
   значения $\delta{P}_{\alpha=0}$ --- используя метод параметризации.
   Величины сдвигов и их ошибки измерены в единицах МэВ/$c$.}
  \label{shifting.relative shifts and their errors.table}
  \centering
  \begin{tabular}{|c||c|c||c|c|}
  \hline
  T$_n$, ГэВ &  $\delta{P}_{2G}$  &  $\sigma(\delta{P}_{2G})$  &  $\delta{P}_{\alpha=0}$  &  $\sigma(\delta{P})$ \\ \hline
  0.55  &  $\quad 10.0\quad$  &  $\quad 3.7\quad$  &  $\quad 9.24\quad$  &  $\quad 3.82\quad$ \\ \hline
  0.8   &  $ 5.7$  &  $2.4$  &  6.42  &  2.51  \\ \hline
  1.0   &  $ 9.6$  &  $1.6$  &  8.79  &  1.32  \\ \hline
  1.2   &  $10.5$  &  $1.5$  &  6.45  &  1.28  \\ \hline
  1.4   &  $ 8.1$  &  $4.2$  &  3.96  &  2.45  \\ \hline
  1.7   &  $10.0$  &  $2.5$  &  7.82  &  1.31  \\ \hline
  1.8   &  $11.2$  &  $3.4$  &  6.01  &  3.06  \\ \hline
  2.0   &  $12.7$  &  $4.2$  &  6.57  &  3.10  \\ \hline
  \end{tabular}
  \end{table}
  \noindent
  Наблюдается хорошее согласие двух способов
  определения величины $\delta{P}$.
  Если фон неупругих процессов невелик, например, при T$_n=550$ и 800\,МэВ,
  упругий и квазиуругий спектры хорошо аппроксимируются функцией двойного Гаусса
  и относительный сдвиг определяется почти также,
  как при использовании метода параметризации.
  Если же энергия пучка увеличивается
  или не работает вето-система,
  как это было при T$_n=1.2$\,ГэВ,
  значения сдвигов, полученные с помощью фита функцией \eqref{setup.2gauss-function},
  имеют тенденцию к завышению.
  Поскольку метод параметризации снижает влияние неупругих реакций,
  мы будем считать эти результаты более надёжными.

  Важно отметить, что сдвиг на 6.5\,МэВ/$c$ в спектре импульсов
  даёт почти такое же смещение по энергии.
  Так, при T$_n=800$\,МэВ, имеем:
  \begin{equation}\label{shifting.energy missing due to the shift 6 MeV/c}
    Ed{E} \;=\; Pd{P}
    \qquad\Rightarrow\qquad
    \delta{E}\Bigr|_{\textrm{800\,МэВ}}\approx\;5.4\,\textrm{МэВ}\;,
  \end{equation}
  что в 2.5 раза превышает энергию связи
  ядра дейтерия.

  \section{Обсуждение эффекта}
  Весьма интересна история наблюдения этого эффекта в прошлом,
  о чём впервые упоминается в статье группы из Лос-Аламоса \cite{Bjork}.
  Принцип определения отношения $R_{dp}$ у них был такой же, --- измеряется
  дифференциальное сечение квазиупругой перезарядки на дейтерии
  и сравнивается с упругой перезарядкой на водороде,
  но вместо нейтронного пучка был протонный
  и рассматривалась другая реакции $pd\to n+(pp)$.
  Измерения проходили при двух энергиях 647 и 800\,МэВ.
  В своей работе авторы сообщают о том,
  что спектр вторичных нейтронов квазиупругой реакции
  оказался смещён по шкале энергии на $\delta{E}\sim7$\,МэВ в сторону меньших значений
  относительно среднего значения импульса протонов пучка.
  Точность измерения этой величины составляет $\approx10$\,МэВ,
  что связано с разрешением их спектрометра.
  Может быть поэтому авторы не акцентируют внимание на данном феномене,
  хотя и говорят о согласии с некоторыми расчётами в диссертационной работе,
  которую выполнил D.\,S. Ward в Техасском университете в 1971 г.
  Определённый интерес представляет вторая статья \cite{Bonner} той же группы.
  Там говорится о другом измерении квазиупругого процесса,
  но теперь используется нейтронный пучок
  и рассматривается {\it наша} реакция $nd\to p+ (nn)$.
  Хотя эффект смещения в статье прямо не обсуждается,
  но там легко найти подтверждение, что он действительно имел место.
  Так, энергия нейтронов в заголовке статьи объявлена равной T$_n=794$\,МэВ,
  но в описании эксперимента говорится,
  что нейтронный пучок получается от перезарядки первичных протонов с энергией T$_p=800$\,МэВ
  на дейтериевой мишени в симметричной квазиупругой реакции $pd\to n(pp)$.
  Нетрудно посчитать, что потеря по шкале импульсов оказывается порядка тех же $7$\,МэВ/$c$.
  В то же время, в публикации остаётся неясным,
  как была получена эта оценка энергии нейтронов.
  Если использовался спектр вторичных протонов
  исследуемой реакции $nd\to p(nn)$,
  то и сам эффект смещения должен был проявиться дважды,
  то есть центральное значение энергии протонов
  при рассеянии под нулём оказалось бы порядка T$_p\sim788$\,МэВ,
  но авторы этих данных не приводят,
  а по изображению спектра на рис.\,1 в их статье \cite{Bonner}
  точное положение центра квазиупругого пика определить нельзя.

   Наша группа впервые обратила внимание на этот феномен после сеанса в декабре 2005 г.,
   когда одна из точек измерялась при энергии 800\,МэВ.
   Условия набора были почти идеальны: большая статистика,
   вклады неупругих реакций и примесь дейтронов полностью исключены 
   из протонных спектров работой систем ДОМ и TOF.
   Аппроксимация обоих пиков определила координаты их центров с такой точностью,
   что сдвиг $\delta{P}\approx6$\,МэВ/$c$ невозможно было игнорировать.
   Кроме того, периодическая смена мишени ($\textrm{CH}_2\to\textrm{CD}_2\to\textrm{C}$) каждые 3\,ч
   исключала систематическую ошибку,
   связанную с точностью установки значений величины поля ускорителя,
   т.е. энергии пучка первичных дейтронов.
   После этого были изучены спектры протонов при других более высоких энергиях.
   Тогда же пришла идея метода определения относительного сдвига
   по правым половинам упругого и квазиупругого спектров.
   Профессор Леонид Николаевич Струнов
   в качестве подтверждения наблюдаемого эффекта привёл две вышеупомянутые статьи.
   Тем не менее, истинная причина смещения оставалась под вопросом.
   Простое указание на то, что реакция $nd\to p(nn)$ не совсем упругая,
   не содержит конкретного объяснения,
   почему потеря \eqref{shifting.energy missing due to the shift 6 MeV/c}
   оказывается намного больше энергии связи нуклонов дейтрона.
   В начале 2006 г. мы обратились за советом к профессору Франсуа Легару,
   и он высказал несколько простых соображений по этому поводу.
   Главную роль здесь играет кинетическая энергия нуклонов внутри ядра,
   и эту долю МэВ необходимо компенсировать,
   чтобы нуклоны стали свободными.

  \subsection{Сдвиг $\delta{P}$ и законы сохранения}
  Трудность вычисления части энергии $\delta{E}=E_n-E_p$, потерянной
  во время квазиупругой реакции $nd\to p(nn)$,
  связана с невозможностью взять табличную массу нейтронной пары,
  поскольку такой частицы просто не существует (в совр. базе ядерных данных).
  Если бы перезарядка нейтрона происходила, например, на гелии $^3$He
  и в результате получался тритий $^3$H, таких вопросов, конечно, не возникало.
  Достаточно использовать законы сохранения энергии и импульса:
  \begin{eqnarray}
   & \sqrt{\phantom{\big|}\!\!m^2_n+{P^2_n}} + m^{\phantom{1}}_{\,^3\textrm{He}} \;=\;
     \sqrt{\phantom{\big|}\!\!m^2_p+{P^2_p}} +
     \sqrt{\phantom{\big|}\!\!m^2_{\,^3\textrm{H}}+{q^2}}\;,
     \label{shifting.charge-exchange helium3 to tritium and energy law} \\ [2mm]
   & \vec{q} = \vec{P_n}-\vec{P_p} \quad\Rightarrow\quad
     q^2 = \delta^2{P} + 4P_n(P_n-\delta{P})\,\sin^2{\dfrac{\theta}{2}}\;\,.
     \label{shifting.charge-exchange and momentum law}
  \end{eqnarray}
  Здесь $\theta$ --- угол рассеяния.
  Величина $\delta{P}=P_n-P_p$ обозначает потерю импульса
  и находится решением уравнения
  \eqref{shifting.charge-exchange helium3 to tritium and energy law}.
  Возвращаясь к нашему случаю, необходимо убрать лишний протон-спектатор:
  \begin{equation}\label{shifting.charge-exchange deutron to dineutron and energy law}
    \sqrt{\phantom{\big|}\!\!m^2_n+{P^2_n}} + m_d \;=\;
    \sqrt{\phantom{\big|}\!\!m^2_p+{P^2_p}} +
    \sqrt{\phantom{\big|}\!\!m^2_{nn}+{q^2}}\;.
  \end{equation}
  Масса $m_{nn}$ это энергия двух нейтронов в их собственной системе покоя.
  Например, если предположить, что рождается нейтральное ядро
  с массой $2m_n\approx1879$\,МэВ/$c^2$,
  то потеря энергии $\delta{E}$ составит $\varepsilon_{\textrm{св}}\approx2.23$\,МэВ
  и величина $\delta{P}$ окажется на уровне $2.5$\,МэВ/$c$
  (рис.~\ref{shifting.shifting experimental data and explanation}),
  чего в экспериментальных данных не наблюдается.
  Тем не менее, саму возможность образования динейтрона отрицать нельзя.
  Согласно оценкам авторов работы \cite{Baz-Goldansky-Zeldovich}
  вероятность такой реакции в 1000 раз меньше,
  чем вероятность процесса упругой $n\to p$ перезарядки.
  Поскольку нейтронный пучок имеет дисперсию $\sigma{P}/P\approx3.3$\,\%,
  различить какой-то особенный вклад реакции с образованием динейтрона
  в спектре импульсов вторичных протонов
  крайне сложно: его будет просто не видно на фоне основной реакции $nd\to p(nn)$.
  Наоборот, по совету Ф.\,Легара
  значение массы $m_{nn}$ следует взять больше $2m_n$ и
  представить в виде суммы:
  \begin{subequations}\label{shifting.mass of dineutron}
  \begin{equation}\label{shifting.mass of dineutron by the Lehar}
    m_{nn} = 2m_n+\varepsilon_{\textrm{св}}\;,
  \end{equation}
  где величина $\varepsilon_{\textrm{св}}\approx2.23$\,МэВ
  выполняет роль кинетической энергии в системе центра масс двух нейтронов.
  Кроме того, 
  мы можем считать, что сразу в момент перезарядки $d\to nn$
  каждый из нейтронов имеет тот же Ферми-импульс $P_F$,
  которым обладали нуклоны внутри ядра дейтерия. Поэтому:
  \begin{equation}\label{shifting.mass of dineutron by the Strunov aspirant}
    m_{nn} = 2\sqrt{m^2_n+P^2_F}\;.
  \end{equation}
  \end{subequations}
  Эти два определения \eqref{shifting.mass of dineutron} эквивалентны,
  если взять $P_F=\sqrt{m_n\varepsilon_{\textrm{св}}}\approx45.7$\,МэВ/$c$.
  Подстановка $m_{nn}$ в уравнение
  \eqref{shifting.charge-exchange deutron to dineutron and energy law}
  позволяет определить абсолютный сдвиг\footnote{Чтобы найти
  величину его смещения относительно упругого,
  для реакции перезарядки $np\to pn$ решается аналогичное уравнение:
  \begin{equation}\label{shifting.charge-exchange neutron to proton and energy law}
    \sqrt{\phantom{\big|}\!\!m^2_n+{P^2_n}} + m_p \;=\;
    \sqrt{\phantom{\big|}\!\!m^2_p+{P^2_p}} +
    \sqrt{\phantom{\big|}\!\!m^2_n+{q^2}}\;,
  \end{equation}
  из которого вычисляется сдвиг $\delta{P}_p$\,,
  отличный от нуля даже при нулевом значении угла $\theta$,
  что связано с разностью масс нейтрона и протона. Тогда:
  \begin{equation}\label{shifting.relative shift using formula}
    \delta{P}\;=\;\delta{P}_{p\,'}\;-\;\delta{P}_p\;.
  \end{equation}}
  $\delta{P}_{p\,'}$ квази-упругого пика.

  \subsection{Интерпретация в рамках модели дейтрона}
  Величина Ферми-импульса $P_F$, который достаётся частице в момент развала ядра,
  имеет случайный характер. Это распределение
  можно описать, например, функцией Хюльтена
  (прил.~\ref{Appendix.Hulthen expression},
  форм.~\ref{appendix.Hulthen expression},~\ref{appendix.Hulthen expression momentum representation}
  и рис.~\ref{appendix.Hulthen expression momentum representation and its probability} слева).
  При этом среднее значение $\overline{P}_F$ составляет $\approx86$\,МэВ/$c$.
  Решение Хюльтена хорошо согласуется с экспериментальными данными
  \cite{Sitnik,Sitnik2} по развалу дейтрона
  (прил.~\ref{Appendix.Hulthen expression},
  рис.~\ref{appendix.Hulthen expression momentum representation and its probability} справа),
  где  $\overline{P}_F$ определяется на уровне $80$\,МэВ/$c$.
  Получено также согласие между этим решением
  и данными нашего эксперимента
  (прил.~\ref{Appendix.neutron momentum distribution},
  рис.~\ref{appendix.neutron momentum distribution at 1.0 GeV}).

  При изложении этого подхода мы неоднократно сталкивались с таким мнением,
  что за счёт Ферми-движения спектр квазиупругой реакции не только сдвинется,
  но весьма существенно изменит и свою форму
  --- он должен размыться в сторону меньших значений импульса.
  Чтобы проверить это предположение,
  было разыграно Монте-Карло и для каждого события уравнение
  \eqref{shifting.charge-exchange deutron to dineutron and energy law}
  решалось отдельно при заданных значениях $P_n$ и $P_F$.
  Для большей точности было учтено,
  что относительное движение между нейтронами
  изменяется на величину переданного импульса $\vec{q}$,
  поэтому в системе центра масс одному из нейтронов,
  который в $nd$-взаимодействии является частицей отдачи, сообщается импульс $\vec{q}/2$,
  а спектарному нейтрону следует прибавить $-\vec{q}/2$.
  Это немного меняет определение инвариантной массы и полной энергии
  двух нейтронов:
  \begin{equation}\label{shifting.mass of dineutron with q-corretion}
    m_{nn} = 2\sqrt{m^2_n+\left(\vec{P}_F+\frac{\vec{q}}{2}\right)^2}\;,\quad
    E_{nn} = \sqrt{4m^2_n+4\left(\vec{P}_F+\frac{\vec{q}}{2}\right)^2+q^2}\;.
  \end{equation}

\begin{figure}[!ht] \vspace {-5mm}
  \centering
  \scalebox{.25}{\includegraphics{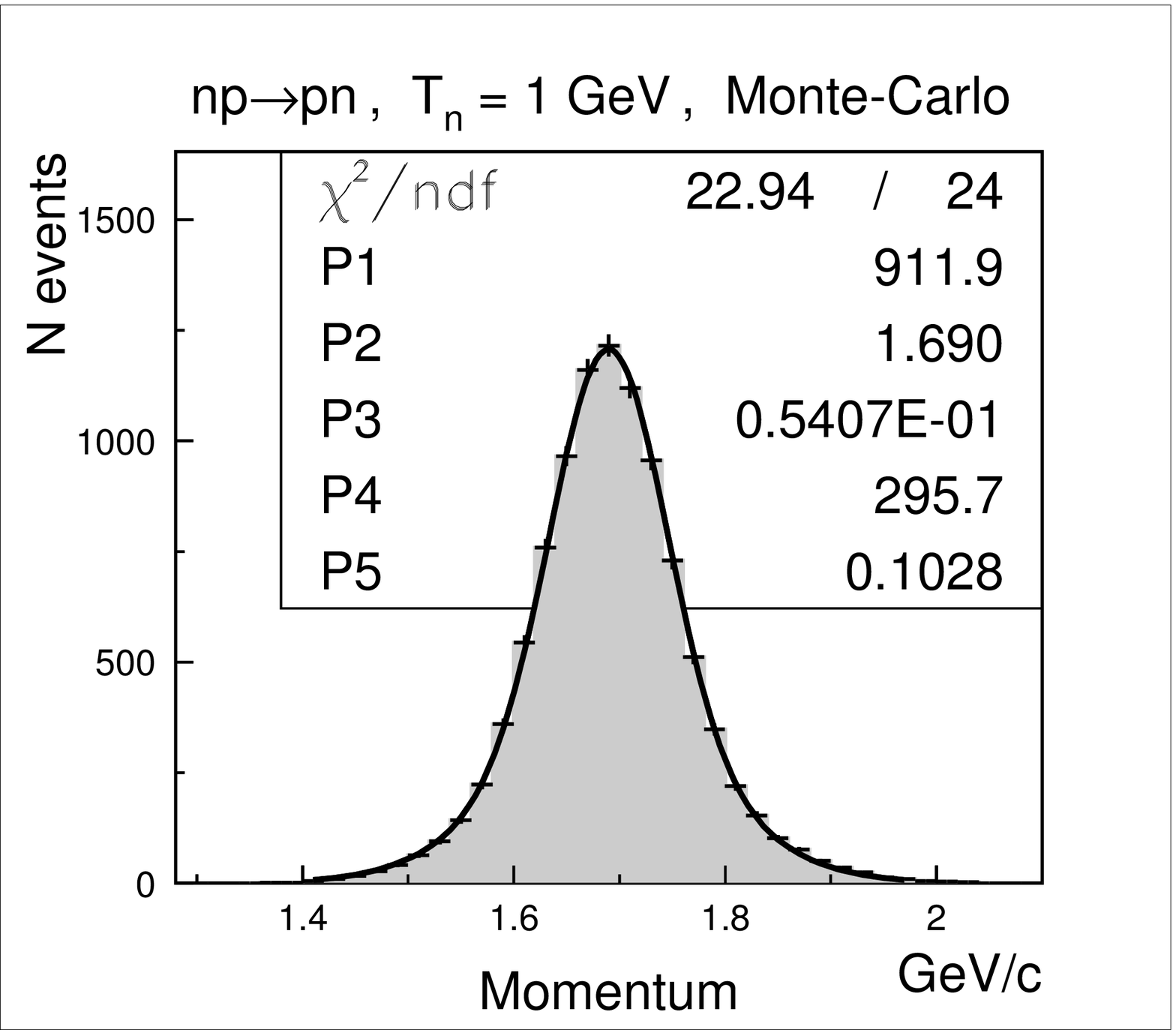}}\quad
  \scalebox{.25}{\includegraphics{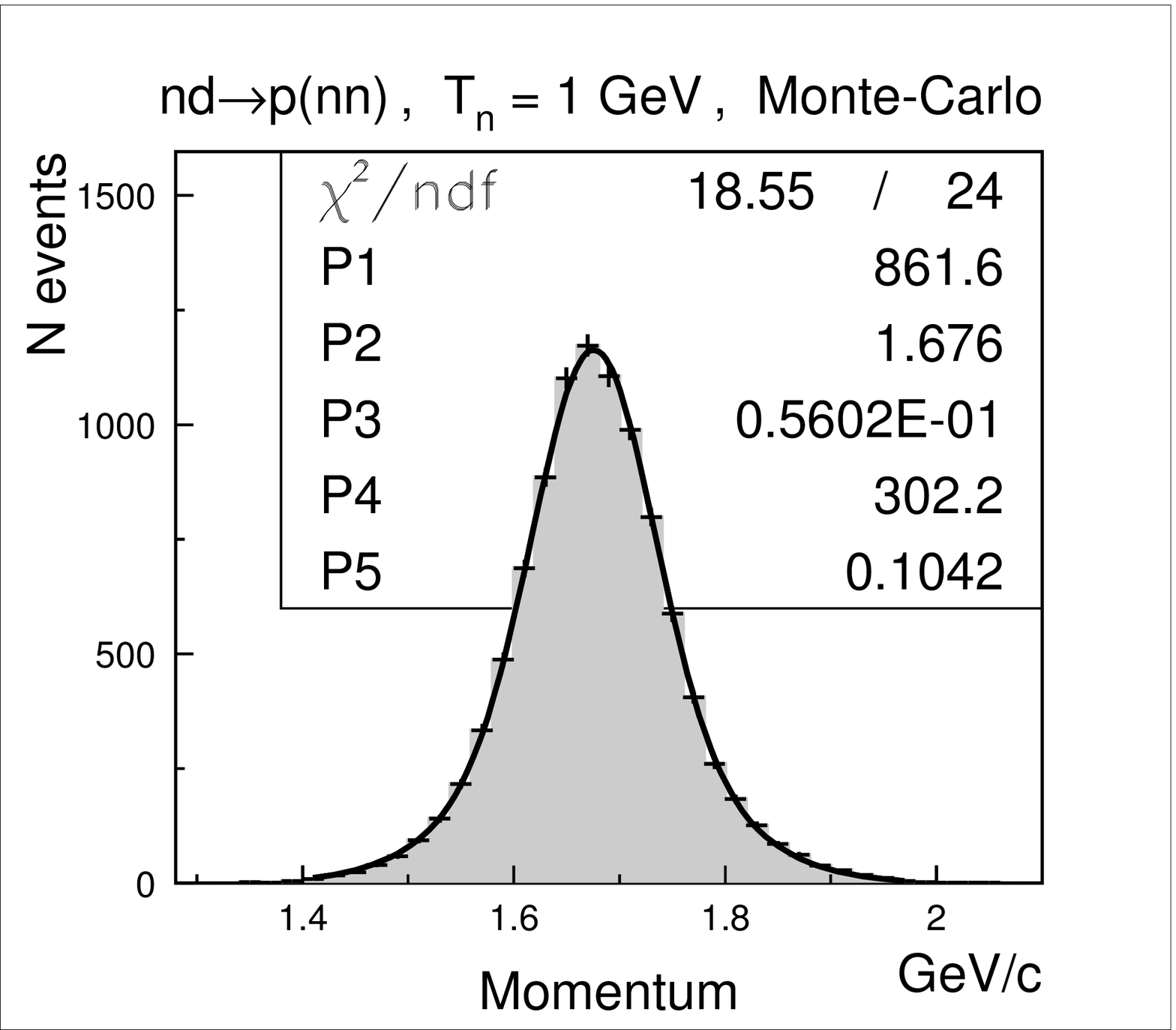}}
  \caption{\footnotesize {Слева спектр импульсов протонов реакции
  $np\to pn$ перезарядки под нулём градусов при энергии T$_n$\,=\,1.0\,ГэВ,
  полученный методом Монте-Карло.
  Справа представлен спектр протонов квазиупругой перезарядки $nd\to p(nn)$,
  где конечное состояние двух нейтронов задано распределением Хюльтена
  (прил.~\ref{Appendix.Hulthen expression},
  форм.~\ref{appendix.Hulthen expression. Estimations of the mean and rms of nuclon momentum}).
  Оба спектра аппроксимируются функцией двойного Гаусса \eqref{setup.2gauss-function}.
  Параметры фита определены также, как на рис. \ref{shifting.respectively shift at 0.8 GeV}.
  Сдвиг между спектрами
  составляет $\delta P=\textrm{p}_2(\textrm{H}_2)-\textrm{p}_2(\textrm{D}_2)=14$\,МэВ/$c$.}}
  \label{shifting.relative shift at 1.0 GeV using Monte-Carlo}
  \end{figure}

  Направление Ферми-импульса $\vec{P}_F$
  полагалось сферически симметричным относительно вектора $\vec{q}$.
  Монте-Карло расчёты были проведены по всему диапазону энергий T$_n=0.5\div2.0$\,ГэВ.
  На рис. \ref{shifting.relative shift at 1.0 GeV using Monte-Carlo}
  представлены спектры реакций $np\to pn$ и $nd\to p(nn)$ при T$_n=1.0$\,ГэВ.
  Распределения имеют почти одинаковую форму,
  то есть никакого существенного размытия
  не происходит\footnote{В обратной кинематике,
  когда протон или другая частица покоится,
  а дейтрон налетает с последующим развалом на два нуклона,
  их Ферми-импульсы складываются с продольными импульсами,
  что приводит к расширению спектра,
  как это наблюдается в импульсном распределении вторичных нейтронов
  (прил.~\ref{Appendix.neutron momentum distribution},
  рис.~\ref{appendix.neutron momentum distribution at 1.0 GeV}).}.
  Спектр квазиупругого процесса $nd\to p(nn)$
  лишь на $2-3$\,МэВ/$c$ шире
  упругого пика $np\to pn$.
  Величина сдвига $\delta{P}$ здесь определяется также,
  как и в экспериментальных данных, по разности между координатами центров пиков:
  \begin{equation}\label{shifting.relative shift using spectra}
    \delta{P}\;=\;\overline{P}_p\;-\;\overline{P}_{p\,'}\;.
  \end{equation}
  \indent
  Проведённые вычисления также показали, что Хюльтен даёт в два раза
  большее значение $\delta{P}$ при всех энергиях T$_n=0.5\div2.0$\,ГэВ,
  нежели реальная величина этого эффекта
  (таб.~\ref{shifting.Monte-Carlo relative shifts and their errors.table}).
  Это расхождение говорит о необходимости другой волновой функции
  двух нейтронов в реакции $nd\to p(nn)$.
  Решение Хюльтена применимо только в том случае,
  если спины нуклонов дейтрона параллельны,
  т.е. когда дейтрон распадается в электромагнитном канале $\gamma+d\to n+p$
  и Ферми-импульсы протона и нейтрона сохраняются
  за счёт сильной связи в триплетном состоянии.
  В нашем случае
  следует говорить о возникновении некоторой системы из двух нейтронов
  с противоположными спинами в сравнительно слабом синглетном состоянии.
  Даже упрощенный подход \eqref{shifting.mass of dineutron by the Strunov aspirant}
  с подстановкой $P_F=45.7$\,МэВ/$c$
  находится в лучшем согласии с экспериментальными данными
  (рис.~\ref{shifting.shifting experimental data and explanation}),
  чем более тщательные расчёты по Хюльтену.
  Измеренные нами значения сдвигов $\delta{P}$ 
  были аппроксимированы,
  используя формулы (\ref{shifting.charge-exchange and momentum law},
  \ref{shifting.charge-exchange deutron to dineutron and energy law},
  \ref{shifting.mass of dineutron with q-corretion}),
  что определило оптимальный Ферми-импульс $P_F$
  на уровне $57.9\pm4.8$\,МэВ/$c$.

  \subsection{Формальный переход $^3S_1\to\,^1S_0$}\label{chapter-Shift, transition 3s1-1s0}
  Чтобы упростить поиск решения,
  величины относительных сдвигов $\delta{P}$
  (таб.~\ref{shifting.relative shifts and their errors.table})
  переведены в шкалу инвариантной массы $nn$-системы
  (таб.~\ref{shifting.mass of nn-pair experimental data.table})
  по формуле \eqref{shifting.charge-exchange deutron to dineutron and energy law}.
  Пренебрегая зависимостью от энергии пучка нейтронов,
  8 значений были аппроксимированы константой
  (рис.~\ref{shifting.mass of nn-pair experimental data}),
  что привело к результату: $\overline{m}_{nn}=1882.7\pm0.6$\,МэВ/$c^2$.
  В основном канале реакции $nd\to p(nn)$
  при рассеянии вторичных протонов вблизи нуля
  $nn$-система обладает избыточной энергией
  $\overline{\varepsilon}_{nn}=3.5\pm0.6$\,МэВ.
 \begin{table}[!ht]
 \parbox{0.47\textwidth}{
 \centering
 \vspace {-1mm}
 \small
 \begin{tabular}{|c|c|}
 \hline
 T$_n$, ГэВ  &  $\;\,\overline{m}_{nn}$, MэВ/$c^2$
 \\ \hline
    0.55   &  1883.5  $\pm$  3.0  \\ \hline
    0.8    &  1881.7  $\pm$  2.1  \\ \hline
    1.0    &  1883.9  $\pm$  1.2  \\ \hline
    1.2    &  1882.0  $\pm$  1.2  \\ \hline
    1.4    &  1879.8  $\pm$  2.3  \\ \hline
    1.7    &  1883.4  $\pm$  1.2  \\ \hline
    1.8    &  1881.7  $\pm$  2.9  \\ \hline
    2.0    &  1882.2  $\pm$  2.9  \\ \hline
 \end{tabular}
 \vspace {2mm}
 \captionof{table}{\footnotesize Средние значения
  инвариантной массы $nn$-пары
  в конечном состоянии реакции $nd\to p(nn)$
  при рассеянии протонов под нулём при энергиях T$_n=0.5\div2.0$\,ГэВ.}
 \label{shifting.mass of nn-pair experimental data.table}}\quad
 \begin{minipage}{0.5\textwidth}
 \centering
 \scalebox{.33}{\includegraphics{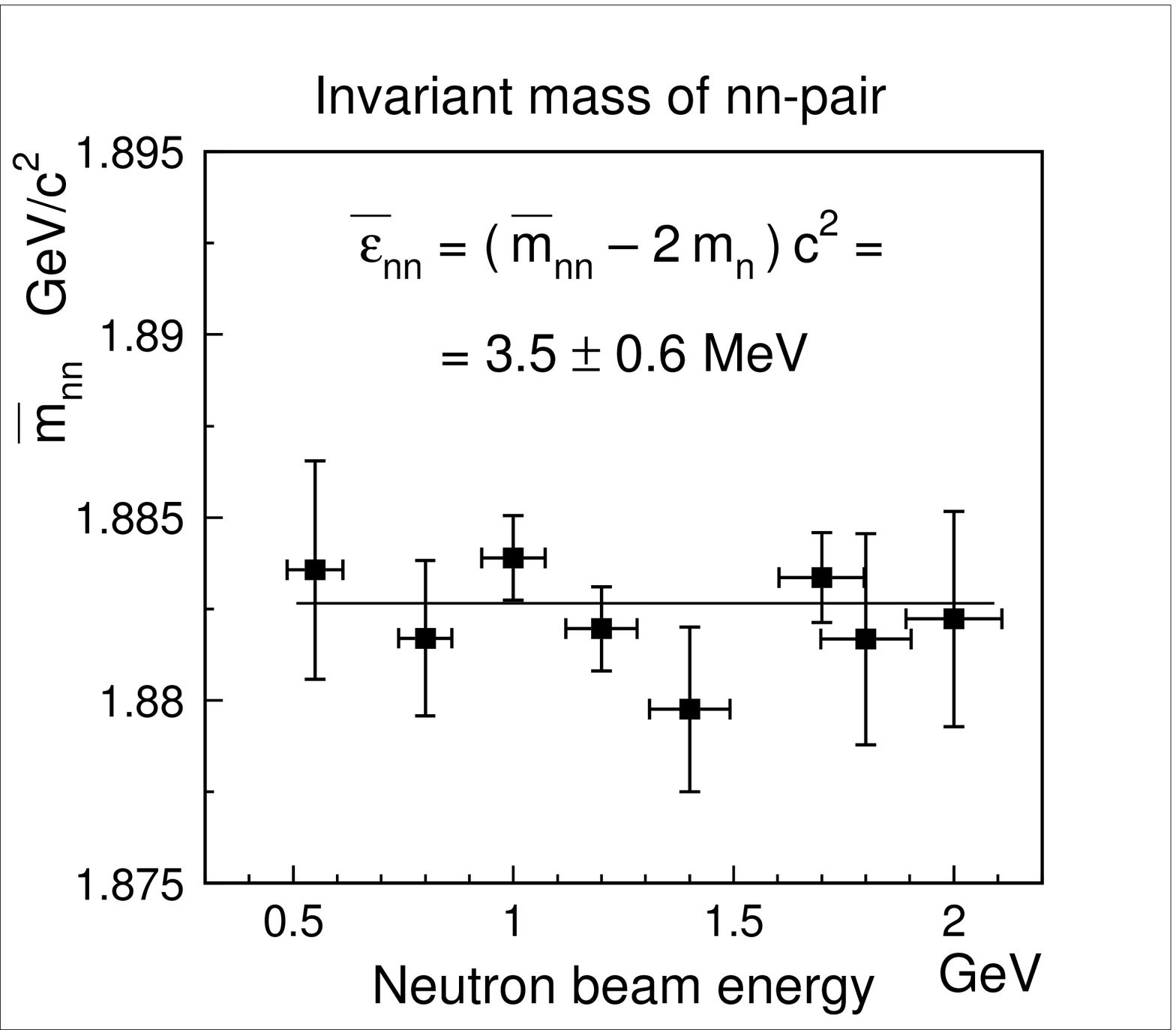}}
 \captionof{figure}{\footnotesize Данные таблицы
 \ref{shifting.mass of nn-pair experimental data.table}
 представлены в графическом виде.
 По восьми независимым измерениям проведён фит константой,
 что определило среднее значение инвариантной массы двух нейтронов
 на уровне $\overline{m}_{nn}=1882.7\pm0.6$\,МэВ/$c^2$.
  }\label{shifting.mass of nn-pair experimental data}
 \end{minipage}
 \end{table}

  Мы будем следовать работе \cite{Lednicky},
  где была рассмотрена аналогичная проблема
  описания системы двух протонов в квазиупругой реакции $dp\to (pp)n$.
  Полагая, что пространственное распределение двух нейтронов
  в момент перезарядки $d\to nn$ соответствует функции Хюльтена 
  (прил.~\ref{Appendix.Hulthen expression}, форм.~\ref{appendix.Hulthen expression}),
  и пренебрегая потенциалом между ними в слабом синглетном состоянии,
  проводим интегрирование с радиальной $s$-волновой асимптотической функцией
  $R_{0}(r, \varphi)$, где $\varphi\equiv\varphi(p)$ ---
  фаза $^1\!S_0$-волны\footnote{
  Значения $\varphi$ являются решениями фазового анализа SP07 \cite{SP07},
  определённого на экспериментальных данных упругого $np$ и $pp$-рассеяний.
  Во втором случае решение добавляет
  кулон-ядерную интерференцию.
  }. Тогда:
  \begin{eqnarray}\label{shifting.1s0-wave calculation}
   \Psi_{nn}(p) &=&
    C_{nn} \int\limits_0^\infty\frac{\,e^{-\alpha r}-e^{-\beta r}}{r}\,
   \frac{\sin({pr}/{\hbar}+\varphi)}{{pr}/{\hbar}}\,4\pi r^2\,dr\;= \nonumber \\ [-2mm]
   &=& C_{nn} \Biggl[\frac{\cos\varphi+\dfrac{\alpha\hbar}{p}\sin\varphi}{(\alpha\hbar)^2+p^2}-
   \frac{\cos\varphi+\dfrac{\beta\hbar}{p}\sin\varphi}{(\beta\hbar)^2+p^2}\Biggr]
    \;.
  \end{eqnarray}
  Константа $C_{nn}$ определяется нормировкой:
  $\int\Psi^2_{nn}\,4\pi p^2\,dp\equiv1$,
  поскольку вклад $d$-волны ($\approx4$\,\%)
  можно считать несущественным,
  на фоне экспериментальной точности ($\approx17$\,\%)
  значения $\overline{\varepsilon}_{nn}$,
  тем более, что при рассеянии вторичных протонов под нулём градусов
  ($q\approx20$\,МэВ/$c$)
  влияние $d$-волны близко к нулевому.

\begin{figure}[!ht]
  \centering
  \scalebox{.3}{\includegraphics{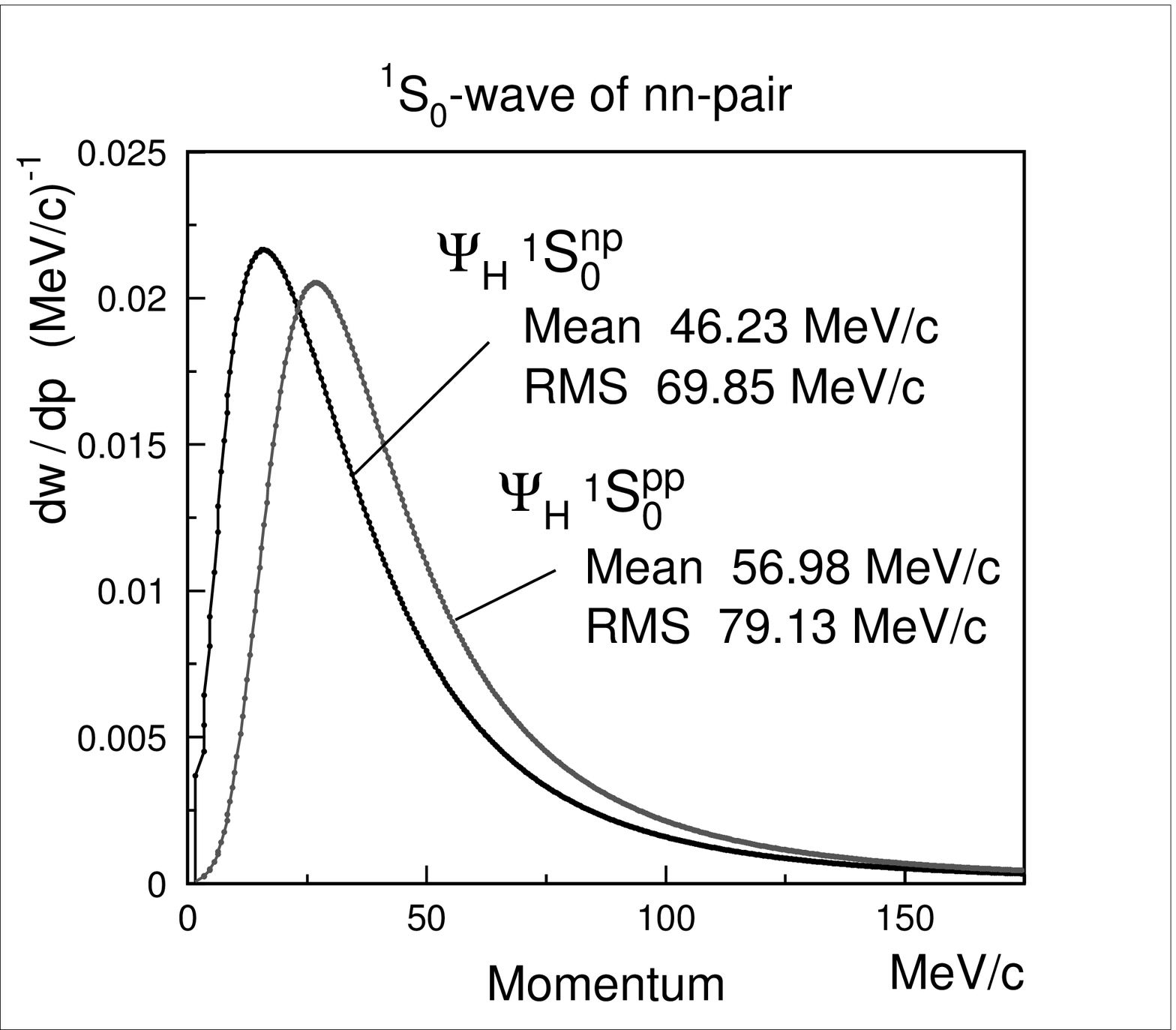}}\quad
  \scalebox{.3}{\includegraphics{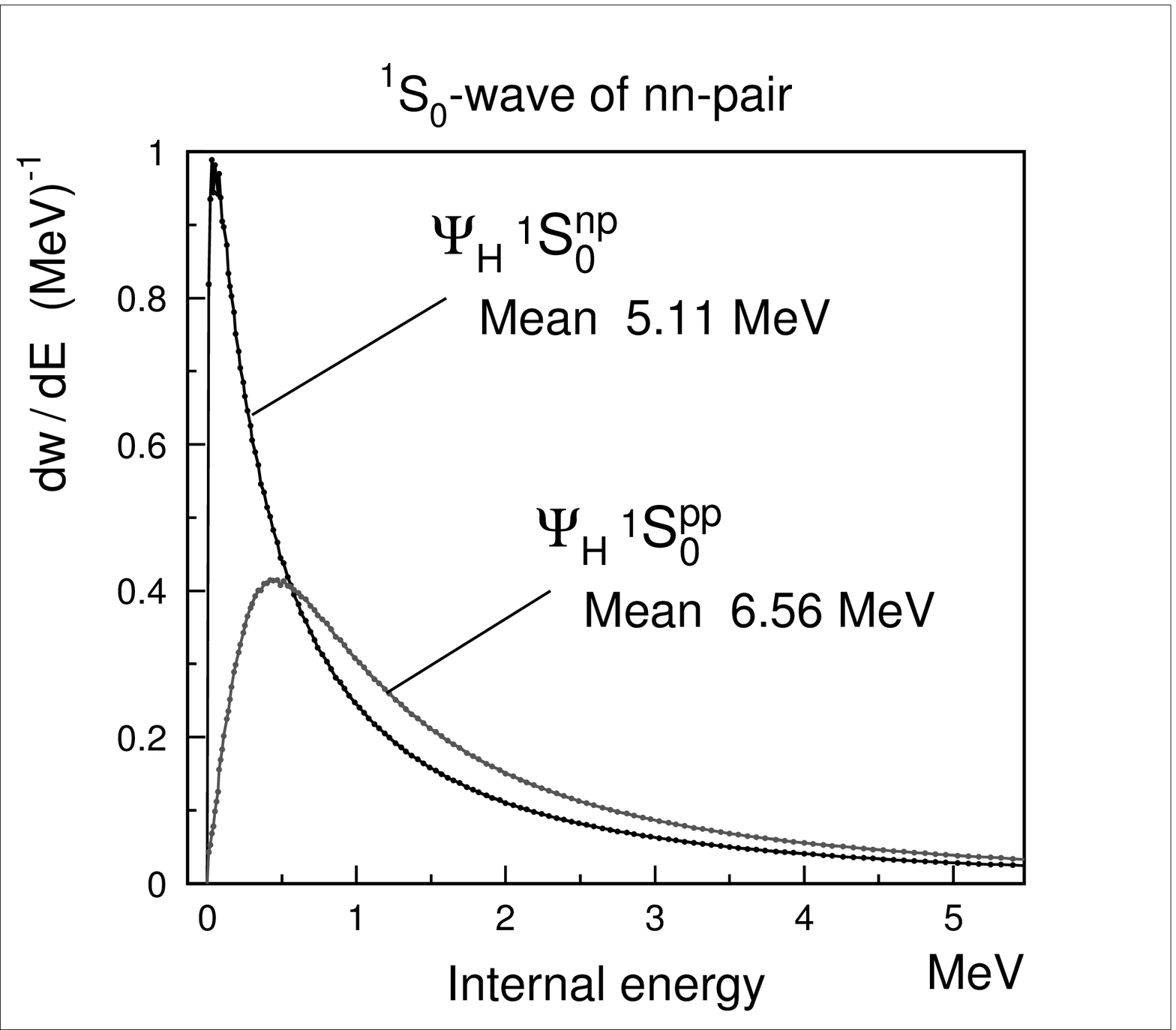}}
  \caption{\footnotesize {Импульсные (слева) и энергетические (справа)
  распределения системы двух медленных нейтронов в конечном состоянии реакции
  $nd\to p(nn)$, рассчитанные по формуле \eqref{shifting.1s0-wave calculation}
  в предположении, что 
  значения фазы $s$-волны $nn$-пары соответствуют
  либо $np$, либо $pp$ упругому рассеянию.
  }}\label{shifting.S-wave of nn-pair}
  \end{figure}

  Расчётные значения $\overline{\varepsilon}_{nn}$
  оказались несколько выше
  3.5\,МэВ  (рис.~\ref{shifting.S-wave of nn-pair}),
  что, очевидно, связано с высокоэнергетичными частями распределений.
  В реальных условиях
  эти вклады уменьшаются вследствие аксептанса спектрометра Дельта-Сигма,
  что было учтено методом Монте-Карло.
  Например, при энергии T$_n=550$\,МэВ вычисления с поправкой на аксептанс
  снижают значения $\overline{\varepsilon}_{nn}$ на 1\,МэВ
  и приводят к оценкам 4.3 и 5.6\,МэВ,
  где фаза $s$-волны берётся из $np$ и $pp$-данных соответственно.

\begin{figure}[!ht]
  \centering
  \scalebox{.35}{\includegraphics{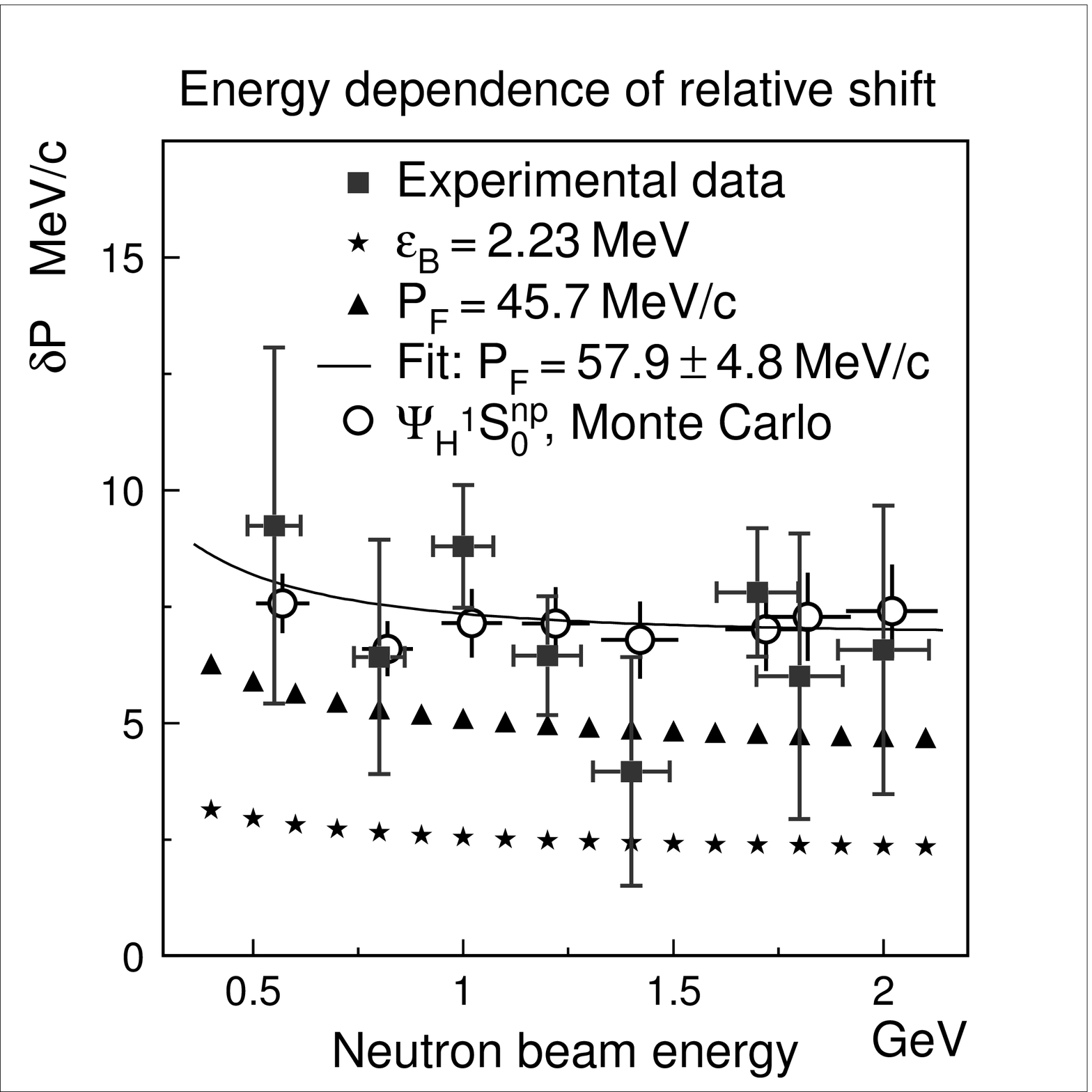}}
  \caption{\footnotesize {Энергетическая зависимость величины относительного сдвига
  между спектрами упругой $np\to pn$ и квазиупругой $nd\to p(nn)$ реакций перезарядки
  в диапазоне энергий T$_n=0.5\div2.0$\,ГэВ.
  Представлены экспериментальные данные, полученные методом параметризации сдвига,
  и 4 варианта расчётных значений: в первом из них смещение $\delta{P}$
  вычисляется в предположении, что 
  энергетические потери связаны только
  с компенсацией энергии связи $\varepsilon=2.23$\,МэВ;
  второй случай соответствует решению алгебраического уравнения
  \eqref{shifting.charge-exchange deutron to dineutron and energy law}
  с постоянным значением импульса Ферми $P_F=45.7$\,МэВ/$c$;
  третий вариант (сплошная линия) представляет результаты фита
  экспериментальных данных функцией $\delta{P}(P_F)$
  (форм.~\ref{shifting.charge-exchange deutron to dineutron and energy law}
  и \ref{shifting.mass of dineutron with q-corretion});
  в четвёртом варианте используется Монте-Карло модель установки
  и внутреннее движение нейтронов задаётся распределением Хюльтена
  с учётом значений фазы $s$-волны рассеяния $nn$-пары
  после смены состояния $^3\!S_1\to\,\!^1\!S_0$
  (п.~\ref{chapter-Shift, transition 3s1-1s0},
  форм.~\ref{shifting.1s0-wave calculation}).
  }}\label{shifting.shifting experimental data and explanation}
  \end{figure}

  Определённые в Монте-Карло расчётах значения сдвига $\delta{P}$
  хорошо согласуются с экспериментальными
  (таб.~\ref{shifting.Monte-Carlo relative shifts and their errors.table}
  и рис.~\ref{shifting.shifting experimental data and explanation}).
  В предположении перехода $nn$-пары в $^1\!S_0$-состояние,
  когда фаза $s$-волны берётся из $np$-данных \cite{SP07},
  среднее значение $\delta{P}$
  в диапазоне энергий T$_n=0.5\div2.0$\,ГэВ
  оказывается на уровне 7\,МэВ/$c$.
  \begin{table}[!ht]
  \caption{\small Значения сдвига $\delta{P}$
   между упругим и квазиупругим пиками в спектрах протонов,
   полученных методом Монте-Карло.
   Импульсное распределение нуклонов задано по функции Хюльтена
   (прил.~\ref{Appendix.Hulthen expression},
   форм.~\ref{appendix.Hulthen expression momentum representation})
   и с поправками на фазовый сдвиг перехода $^3\!S_1\to\,\!^1\!S_0$
   (п.~\ref{chapter-Shift, transition 3s1-1s0},
   форм.~\ref{shifting.1s0-wave calculation}).}
  \label{shifting.Monte-Carlo relative shifts and their errors.table}
  \centering
  \begin{tabular}{|c|c|c|c|}
  \hline
   & \multicolumn{3}{c|}{$\delta{P}$,\;\textrm{МэВ}/$c$} \\ \hline
  T$_n$, ГэВ &  $\qquad\Psi_H\qquad$  &  $\quad\Psi_H\,^1\!S_0^{\,pp}\quad$  &  $\quad\Psi_H\,^1\!S_0^{\,np}\quad$ \\ \hline
  0.55  &  $15.0 \pm 0.9$  &  $9.0 \pm 0.6$  &  $7.6 \pm 0.6$  \\ \hline
  0.8   &  $13.2 \pm 0.8$  &  $7.8 \pm 0.6$  &  $6.6 \pm 0.6$  \\ \hline
  1.0   &  $14.6 \pm 1.1$  &  $8.5 \pm 0.7$  &  $7.1 \pm 0.7$  \\ \hline
  1.2   &  $14.2 \pm 1.1$  &  $8.5 \pm 0.8$  &  $7.1 \pm 0.8$  \\ \hline
  1.4   &  $13.3 \pm 1.2$  &  $8.1 \pm 0.8$  &  $6.8 \pm 0.8$  \\ \hline
  1.7   &  $13.9 \pm 1.3$  &  $8.4 \pm 0.9$  &  $7.0 \pm 0.9$  \\ \hline
  1.8   &  $14.7 \pm 1.4$  &  $8.7 \pm 1.0$  &  $7.3 \pm 0.9$  \\ \hline
  2.0   &  $15.0 \pm 1.4$  &  $8.8 \pm 1.0$  &  $7.4 \pm 1.0$  \\ \hline
  \end{tabular}
  \end{table}

  Второй вариант,
  когда $\varphi$ соответствует упругому $pp$-рассеянию,
  более подходит для реакции $dp\to (pp)n$,
  в которой регистрируются два быстрых протона.
  Примером такого опыта  при энергии T$_d=1.17$\,ГэВ
  с хорошим разрешением протонных импульсных спектров 
  может служить эксперимент на установке ANKE COSY \cite{Epp}.
  При малых значениях переданного импульса $q\in[0,\,100]$\,МэВ/$c$\,
  энергетическая зависимость величины $d\sigma/d\varepsilon_{pp}$,
  где $\varepsilon_{pp}$ --- избыток энергии в $pp$-системе,
  совпадает с полученным нами распределением
  (рис.~\ref{shifting.S-wave of nn-pair},
  $\Psi_H\,^1\!S_0^{\,pp}$).

  \pagebreak

  Следует отметить, что
  фазовый сдвиг $s$-волны в переходе $^3\!S_1\to\,\!^1\!S_0$ является формальным.
  В рамках импульсного приближения,
  в котором происходит вывод формулы Дина \eqref{introduction.Rdp-ratio.formula}
  \cite{Dean-1, Lednicky}, значение $\varphi$ должно равняться нулю.
  Для более строгого описания квазиупругой реакции $nd\to p(nn)$
  необходимо развивать альтернативный способ,
  например, используя уравнение Липмана-Швингера,
  что позволит учесть в процессе $d\to nn$ перезарядки изменение потенциальной энергии
  при переходе нуклонов из триплетного состояния в синглетное.

  Несмотря на то, что решение \eqref{shifting.1s0-wave calculation}
  очень близко к экспериментальным данным,
  остаются нерешёнными некоторые вопросы.
  Например, если $nn$-пара обладает непрерывным спектром
  (рис.~\ref{shifting.S-wave of nn-pair}), 
  девиация инвариантной массы $m_{nn}$ двух нейтронов,
  как говорилось выше, привела бы к расширению на $2-3$\,МэВ/$c$
  импульсного спектра вторичных протонов.
  Однако при энергиях T$_n=550$ и 800\,МэВ этого не наблюдается,
  а при более высоких значениях энергии
  на оценку ширины пика начинают влиять события неупругих реакций,
  которые не удалось подавить вето-системой ДОМ.

\clearpage
\section{Заключение}
 Представлены данные наблюдаемого смещения
 спектра импульсов протонов квазиупругой реакции $nd\to p(nn)$
 относительно спектра упругой реакции $np\to pn$
 в сторону меньших значений при энергиях T$_n=0.5\div2.0$\,ГэВ.
 Величина этого сдвига составляет $\delta{P}=6.5\pm2.5$\,МэВ/$c$.

 Рассмотрен вопрос влияния неупругих процессов,
 связанных с рождением резонанса $\Delta\,(1232)$,
 на координаты центров упругого и квазиупругого пиков,
 определёных методом фита функцией двойного Гаусса.
 Разработан и применён альтернативный способ параметризации сдвига,
 позволяющий находить значение $\delta{P}$,
 используя только правые {\it хорошие} половины импульсных спектров,
 где влияние неупругого фона сведено к минимуму
 вследствие кинематического предела рождения $\Delta$.

 Проведена проверка различных моделей,
 предлагающих описание двух медленных нейтронов в конечном состоянии
 реакции $nd\to p(nn)$.
 Решение Хюльтена
 даёт смещение спектра вторичных протонов на величину $\delta{P}\approx14$\,МэВ/$c$,
 что дважды превышает наблюдаемое значение.
 Если состояние $nn$-пары задаётся Ферми-импульсом $P_F\approx57.9$\,МэВ/$c$,
 вычисления приводят к правильному сдвигу $\delta{P}\approx6.5$\,МэВ/$c$.

 Изменение состояния $^3\!S_1\to\,^1\!S_0$
 с учётом значений фазы $s$-волны от данных упругого $np$ и $pp$-рассеяний
 даёт оценки внутренней энергии $nn$-пары на уровне 5 и 6.6\,МэВ соответственно,
 что близко к величине $\overline{\varepsilon}_{nn}=3.5\pm0.6$\,МэВ,
 определённой экспериментально.
 В рамках аксептанса спектрометра Дельта-Сигма
 этот подход хорошо согласуется
 с наблюдаемыми значениями сдвига $\delta{P}$
 между пиками вторичных протонов
 упругой $np\to pn$ и квазиупругой $nd\to p(nn)$ реакций.

  Для более точного решения вопроса
  о состоянии $nn$-системы
  необходимо повысить качество проведения эксперимента.
  Например, можно уменьшить ошибки наблюдаемых значений сдвига $\delta{P}$,
  если использовать нейтронный пучок c разбросом импульсов $\sigma{P}_n/P_n\approx1.0$\,\%,
  получая его от перезарядки протона на D$_2$-мишени,
  как это делалось в лаборатории Лос-Аламоса \cite{Bonner}.

 \vspace {3mm}
 \thanks{Значительный вклад в это исследование был внесён теми, кого уже нет рядом с нами:
 профессорами Л.\,Н.~Струновым, В.\,Л.~Любошицем и Ф.~Легаром.
 Мы выражаем благодарность научным сотрудникам Объединённого Института Ядерных Исследований
 Ю.\,Н.~Узикову и С.\,С. Шиманскому
 за помощь и теоретические консультации.
 Наш эксперимент был поддержан Российским Фондом Фундаментальных
 Исследований, гранты № 02-02-17129 и № 07-02-01025.}

\appendix

 \section{Функция Хюльтена}\label{Appendix.Hulthen expression}
\begin{figure}[!ht]
  \centering
  \scalebox{.27}{\includegraphics{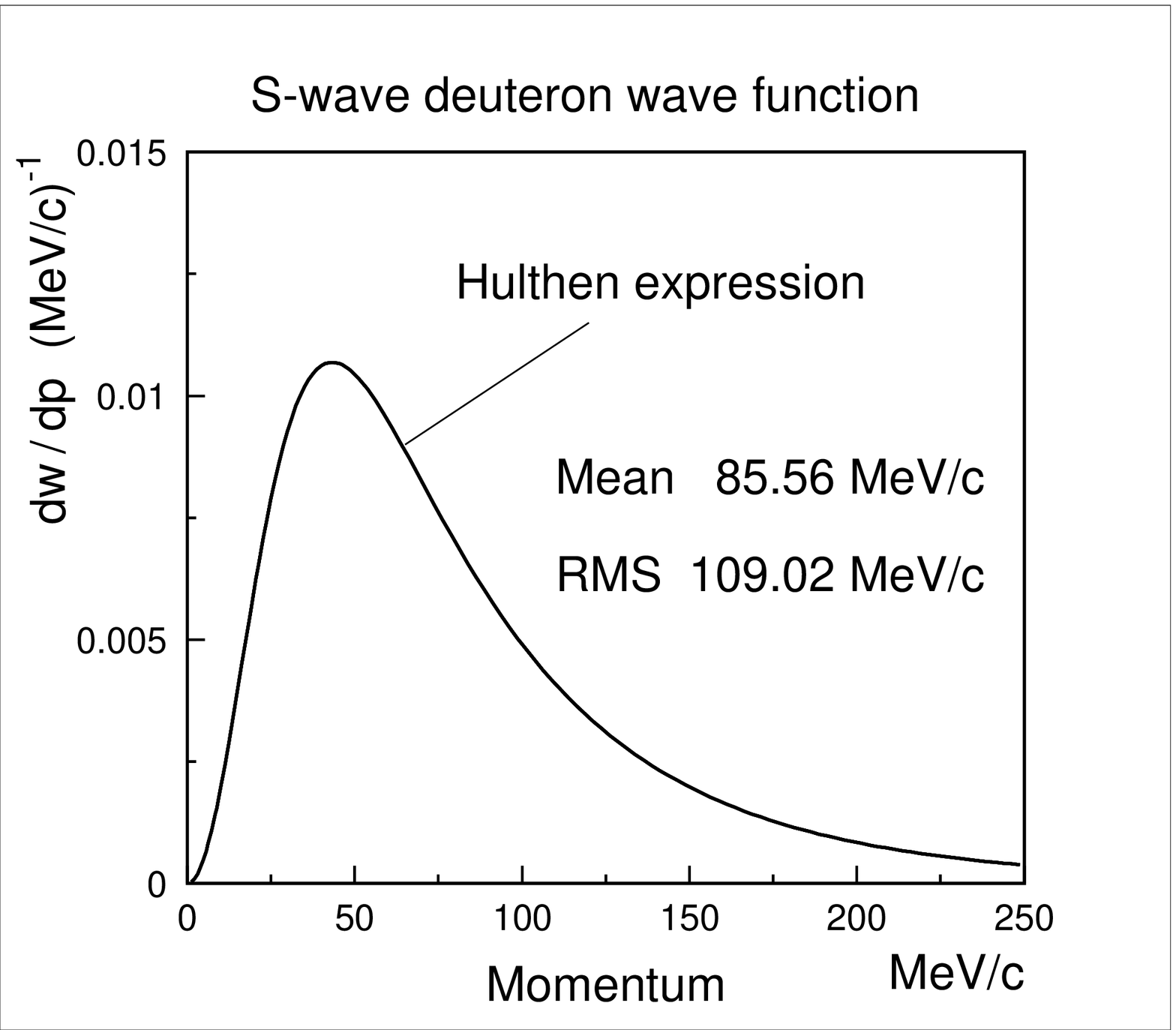}}\quad
  \scalebox{.27}{\includegraphics{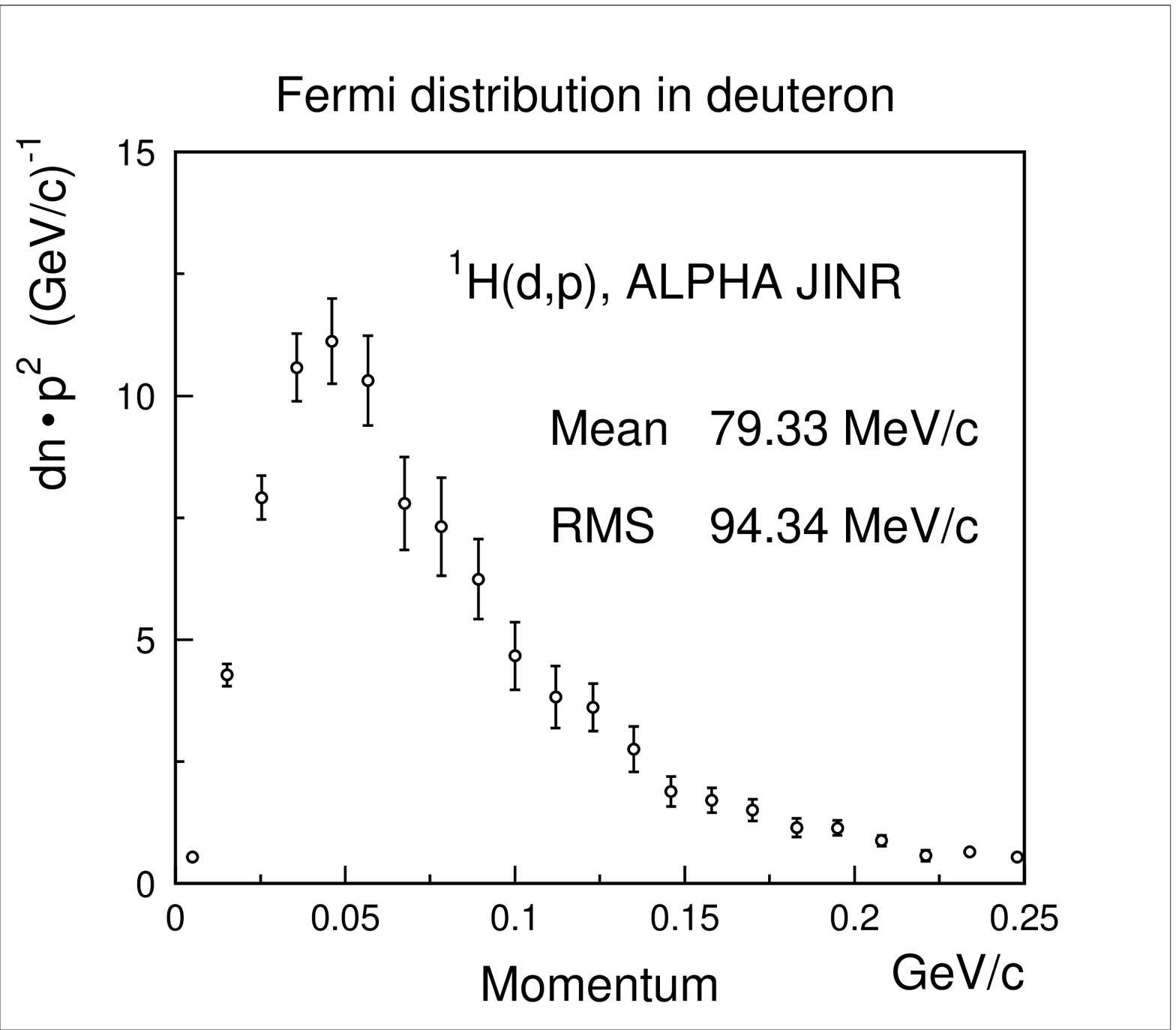}}
  \caption{\footnotesize{Слева плотность вероятности найти нуклоны дейтрона c импульсом $p$.
  Для построения используется формула $dw/dp=\Psi^2_H(p)\,4\pi p^2$,
  где: $\Psi_H(p)$ --- Фурье образ функции Хюльтена \eqref{appendix.Hulthen expression momentum representation}.
  Справа --- экспериментальные данные,
  которые любезно предоставил наш коллега И.\,М.\,Ситник
  по результатам изучения реакции развала дейтрона на установке Альфа \cite{Sitnik,Sitnik2}.
  Согласие эксперимента и теории здесь очевидно подтверждает выбор функции Хюльтена.
  }}\label{appendix.Hulthen expression momentum representation and its probability}
  \end{figure}

  Л.~Хюльтен и М.~Сагавара в 1951 г. в работе \cite{Hulthen}
  для описания пространственного распределения нуклонов дейтрона
  предложили функцию:
   \begin{equation} \label{appendix.Hulthen expression}
    \Psi_H(r)=C\frac{\,e^{-\alpha r}-e^{-\beta r}}{r}\;,
    \quad\textrm{где:}\quad
    C=\sqrt{\frac{\alpha\beta(\alpha+\beta)}{2\pi(\alpha-\beta)^2}}\;,
  \end{equation}
  где: $\alpha=45.7$\,МэВ/$\hbar c$
  и $\beta=260$\,МэВ/$\hbar c$.
  Константа $C$ определяется из условия нормировки: $\int\Psi^2_H(r)\,dV\equiv1$.
  Распределение нуклонов по импульсам
  находится свёрткой функции \eqref{appendix.Hulthen expression}
  с гармоникой $e^{-\tfrac{i}{\hbar}\vec{p}\vec{r}}$:
  \begin{eqnarray}\label{appendix.Hulthen expression momentum representation}
    \Psi_H(p) &=&
    \frac{\sqrt{\hbar\alpha\beta(\alpha+\beta)}}{\pi|\alpha-\beta|}\,
    \left[\frac1{(\alpha\hbar)^2+p^2}-\frac1{(\beta\hbar)^2+p^2}\right]\;.
  \end{eqnarray}
  Вероятность найти нуклоны со значением импульса от $p$ до $p+dp$
  определяется так: $dw(p)=\Psi^2_H(p)\,4\pi p^2dp$
  (рис.~\ref{appendix.Hulthen expression momentum representation and its probability}).
  Среднее значение и среднеквадратичное рассчитаны по формулам:
  \begin{equation}\label{appendix.Hulthen expression. Estimations of the mean and rms of nuclon momentum}
    \left<p\right> = \hbar\,\frac{4\alpha\beta(\alpha+\beta)}{\pi(\alpha-\beta)^2}
    \left(\frac{\alpha^2+\beta^2}{\beta^2-\alpha^2}\ln{\frac\beta\alpha}-1\right)
    \;, \qquad
    \sqrt{\left<p^2\right>} = \hbar\,\sqrt{\alpha\beta}
    \;.
  \end{equation}

  \section{Распределение пучка нейтронов \\ по импульсам}\label{Appendix.neutron momentum distribution}
\begin{figure}[!ht]
  \centering
  \scalebox{.27}{\includegraphics{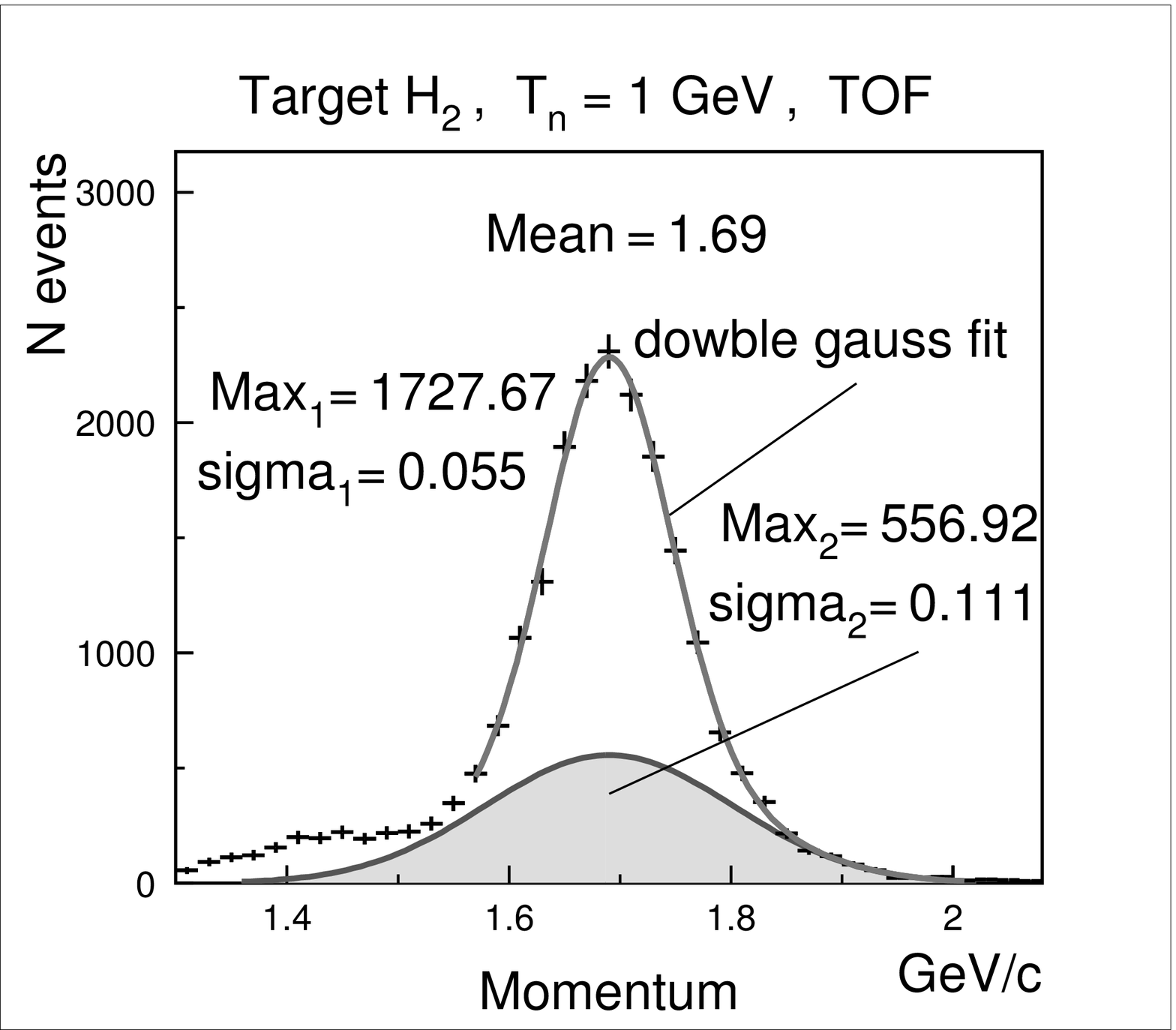}}\quad
  \scalebox{.27}{\includegraphics{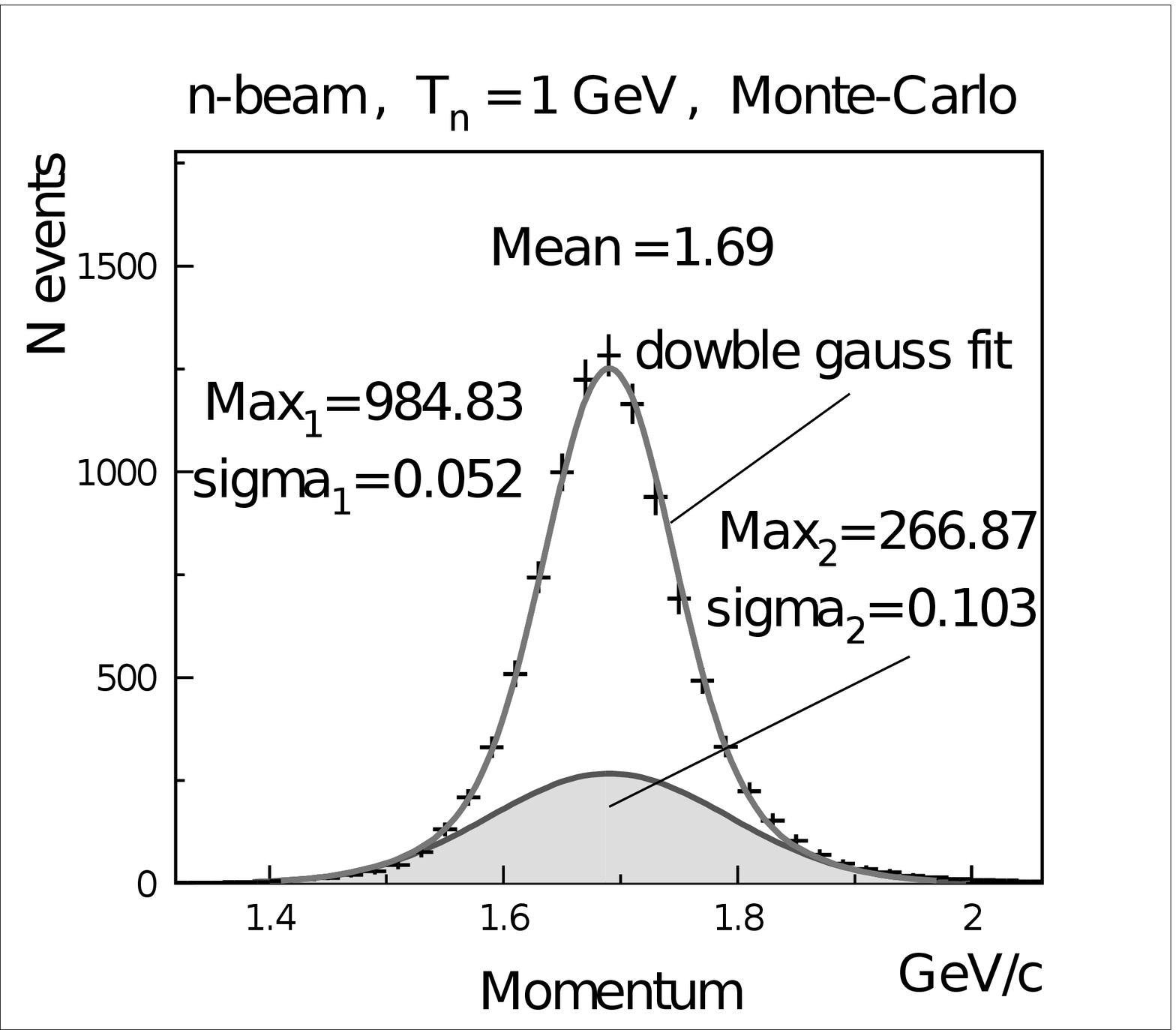}}
  \caption{\footnotesize{Слева спектр импульсов протонов
  $n\to p$ перезарядки под нулём градусов при энергии T$_n=1.0$\,ГэВ.
  Справа представлен моделируемый спектр нейтронов от
  реакции $d\to n+p$ развала первичных дейтронов с энергией T$_d=2.0$\,ГэВ.
  Состояние ядер дейтерия задаётся распределением Хюльтена
  \eqref{appendix.Hulthen expression momentum representation}.
  Оба спектра аппроксимируются функцией
  двойного Гаусса \eqref{setup.2gauss-function}.
  }}
  \label{appendix.neutron momentum distribution at 1.0 GeV}
  \end{figure}

  Нейтронный пучок организуется от развала первичных дейтронов
  на бериллиевой или углеродной мишени, после чего коллиматор
  ограничивает его направление в пределах углов $\sigma\theta\approx2$\,мрад.
  Разброс импульсов пучка нейтронов
  определяется состоянием ядер дейтерия в момент развала.
  В процессе упругой $np\to pn$ перезарядки под нулём
  происходит полная передача импульса $P_p\approx P_n$,
  поэтому спектр вторичных протонов повторяет спектр нейтронов.
  В качестве примера на рис. \ref{appendix.neutron momentum distribution at 1.0 GeV}
  слева представлен спектр импульсов протонов
  $n\to p$ перезарядки под нулём при энергии T$_n=1.0$\,ГэВ.
  Справа показан спектр нейтронов от
  реакции развала первичных дейтронов с энергией T$_d=2.0$\,ГэВ,
  полученный в рамках Монте-Карло модели установки Дельта-Сигма
  (рис.~\ref{setup.spectrometer}).
  Внутреннее движение нуклонов в ядре дейтерия
  задаётся по функции Хюльтена в импульсном представлении
  (прил.~\ref{Appendix.Hulthen expression},
  форм.~\ref{appendix.Hulthen expression momentum representation}).
  Спектры аппроксимируются функцией
  двойного Гаусса \eqref{setup.2gauss-function}.
  Легко видеть, что они обладают почти одинаковой формой,
  что говорит о согласии экспериментальных данных с расчётными.
  Подобное соответствие наблюдается
  во всём диапазоне энергий T$_n=0.5\div2.0$\,ГэВ.


\bibliographystyle{my-ieeetr}
\bibliography{./bib/shindin_dubna_bib/shindin_ru}

\begin{thebibliography}{10}

\bibitem{Lapidus-ppt-method}
С.~М. Биленький, Л.~И. Лапидус, Р.~М. Рындин, ``Поляризованная протонная
  мишень в опытах с частицами высоких энергий,'' {\em УСП. ФИЗ. НАУК}
  \textbf{84}, pp.~243--301, Октябрь 1964.

\bibitem{Lehar-ppt-method}
F.~Lehar {\em et~al.}, ``The movable polarized target as a basic equipment for
  high energy spin physics experiments at the jinr-dubna accelerator complex,''
  {\em Nucl. Instr. Meth. A} \textbf{356}, pp.~58--61, 1995.

\bibitem{Sharov-dsl-measurement}
V.~I. Sharov, L.~N. Strunov, {\em et~al.}, ``Measurements of the total cross
  section difference $\Delta\sigma_{\textrm L}$(np) at 1.59, 1.79 and 2.20
  GeV,'' {\em Eur. Phys. J.} \textbf{C}, no.~13, pp.~255--265, 2000.

\bibitem{Sharov-dsl-measurements}
V.~I. Sharov, L.~N. Strunov, {\em et~al.}, ``Measurements of the total cross
  section difference $\Delta\sigma_{\textrm L}$(np) at 1.39, 1.69, 1.89 and
  1.99 GeV,'' {\em Physics of Atomic Nuclei} \textbf{68}, no.~11,
  pp.~1796--1811, 2005.

\bibitem{Dean-1}
N.~W. Dean, ``Symmetrization effect in spectator momentum distribution,'' {\em
  Phys. Rev. D} \textbf{5}, no.~7, pp.~1661--1666, 1972.

\bibitem{Shindin-Flip-PEPAN}
Р.~А. Шиндин, Д.~К. Гурьев, А.~А. Морозов, {\em и~др.}, ``Разделение
  дифференциального сечения перезарядки $np\to pn$ на flip и non-flip части при
  энергии Т$_n=0.5-2.0$ ГэВ,'' {\em Письма в ЭЧАЯ} \textbf{8}, no.~2 (165),
  pp.~157--168, 2011.

\bibitem{Migdal}
А.~Б. Мигдал, ``Теория ядерных реакций с образованием медленных частиц,'' {\em
  ЖЭТФ} \textbf{28}, pp.~3--9, 1955.
\newblock Доложено на теоретическом семинаре в Институте физических проблем в
  октябре 1950 г.

\bibitem{Baz-Goldansky-Zeldovich}
А.~И. Базь, В.~И. Гольданский, Я.~Б. Зельдович, ``Систематика легчайших
  ядер,'' {\em УСП. ФИЗ. НАУК} \textbf{85}, pp.~445--483, Март 1965.

\bibitem{Shindin-DTS}
R.~A. Shindin, E.~V. Chernykh, D.~K. Guriev, {\em et~al.}, ``Veto-system in
  measurement of the elastic (n,p) charge exchange using H$_2$/D$_2$-target at
  energies T$_n=1-2$ GeV,'' {\em Czech. J. Phys.} \textbf{55}, no.~1,
  pp.~A399--A405, 2005.
\newblock Proceedings of International conference SPIN-PRAHA 2004.

\bibitem{Bjork}
C.~W. Bjork, P.~J. Riley, B.~E. Bonner, {\em et~al.}, ``Neutron spectra at
  $0^{\,\circ}$ from p-p and p-d collisions at 647 and 800 MeV incident
  energies,'' {\em Phys. Lett.} \textbf{B}, no.~63, pp.~31--34, 1976.

\bibitem{Bonner}
B.~E. Bonner, J.~E. Simmons, J.~M. Wallace, {\em et~al.}, ``Quasielastic charge
  exchange in $n\,^2\textrm{H}\to pnn$ at 794 MeV,'' {\em Phys. Rev.}
  \textbf{C}, no.~17, pp.~664--670, 1978.

\bibitem{Sitnik}
В.~Г. Аблеев, Л.~Визирева, В.~И. Волков, {\em и~др.}, ``Измерение тензорной
  анализирующей способности реакции С(d,p) с вылетом протонов под нулевым углом
  при импульсе дейтронов 9.1 ГэВ/c,'' {\em Письма в ЖЭТФ} \textbf{47},
  pp.~558--561, 10 июня 1988.

\bibitem{Sitnik2}
I.~Atanasov, I.~M. Sitnik, {\em et~al.}, ``The measurements of the polarization
  transfer coefficient in (d,p) reaction at fixed proton momentum of 4.5 gev/c
  and the deuteron momentum in range 6.0-9.0 GeV/c,'' in {\em 11th Int. Seminar
  on High Energy Phys. Problem} (A.~M. Baldin and V.~V. Burov, eds.), (Dubna),
  p.~443, JINR, September 7-12 1994.

\bibitem{Lednicky}
R.~Lednicky, V.~L. Lyuboshitz, V.~V. Lyuboshitz, ``Spin effects and
  relative momentum spectrum of two protons in deuteron charge-exchange
  breakup,'' in {\em Proceedings of the XVI International Baldin Seminar on
  High Energy Physics Problems}, vol.~1, pp.~199--211, 2003.

\bibitem{SP07}
R.~A. Arndt {\em et~al.}, ``Updated analysis of nn elastic scattering to 3
  GeV,'' {\em Phys. Rev.} \textbf{C}, no.~76, p.~025209, 2007.

\bibitem{Epp}
D.~Chiladze {\em et~al.}, ``The $dp\to ppn$ reaction as a method to study
  neutron-proton charge-exchange amplitudes,'' {\em Eur. Phys. J. A.}
  \textbf{40}, pp.~23--33, 2009.

\bibitem{Hulthen}
L.~Hulth\'{e}n and M.~Sugavara, {\em The Two-Nucleon Problem}, vol.~39,
  pp.~32--33, 76, 92.
\newblock Springer-Verlag, Berlin, 1957.

\end{thebibliography}


\end{document}